\documentclass{article}

\usepackage[english]{babel}
\usepackage[utf8x]{inputenc}
\usepackage[T1]{fontenc}

\usepackage[a4paper,top=3cm,bottom=2cm,left=3cm,right=3cm,marginparwidth=1.75cm]{geometry}

\usepackage{authblk}

\usepackage{algorithm,algpseudocode}
\usepackage{caption}
\usepackage{bm}
\usepackage{amsmath,amssymb}
\usepackage{graphicx}
\usepackage{cleveref}
\usepackage{subfig}
\usepackage{physics}
\usepackage{tabularx}
\usepackage{threeparttable}
\usepackage{placeins}
\usepackage{jabbrv}

\usepackage{booktabs}
\usepackage{arydshln}
\usepackage{siunitx}
\usepackage{mathtools}

\numberwithin{equation}{section}

\Crefname{figure}{Fig.}{Figs.}
\Crefname{equation}{Eq.}{Eqs.}

\graphicspath{{../figures}}

\renewcommand{\bar}[1]{\overline{#1}}

\newcommand{\minv}[1]{#1^{-1}}

\newcommand{\OO}{\ensuremath{\mathcal{O}}}
\newcommand{\PP}{\ensuremath{\mathcal{P}}}

\newcommand{\OOc}{\OO^*}
\newcommand{\PPc}{\PP^*}

\newcommand{\FF}{\bm{\mathcal{F}}}
\newcommand{\SSn}{\bm{\mathcal{S}}}
\newcommand{\SSk}{\SSn_k}
\newcommand{\sOO}{{_{\OO}}}
\newcommand{\sPP}{{_{\PP}}}

\DeclareMathOperator*{\argmin}{argmin}

\renewcommand{\vec}[1]{\bm{#1}}
\newcommand{\mat}[1]{\mathcal{#1}}

\newcommand{\zz}{\vec{z}}
\newcommand{\yy}{\vec{y}}
\newcommand{\hh}{\vec{h}}
\newcommand{\vv}{\vec{v}}
\renewcommand{\aa}{\vec{a}}
\newcommand{\zt}{\tilde{z}}
\newcommand{\zzt}{\vec{\zt}}

\newcommand{\bomega}{\bm{\omega}}
\newcommand{\bpsi}{\bm{\psi}}
\newcommand{\bzeta}{\bm{\zeta}}

\newcommand{\bPsi}{\bm{\Psi}}
\newcommand{\hpsi}{\hat{\psi}}
\newcommand{\hbPsi}{\bm{\hat{\Psi}}}
\newcommand{\hbpsi}{\bm{\hat{\psi}}}
\newcommand{\hbpsic}{\hbpsi^*}
\newcommand{\bLambda}{\bm{\Lambda}}

\newcommand{\LL}{\mathfrak{L}}
\newcommand{\rr}{\vec{r}}
\newcommand{\LLg}{\LL_g}

\newcommand{\ffg}{f_g}
\newcommand{\ffp}{f_p}

\newcommand{\JJ}{\mat{J}}

\newcommand{\II}{\mat{I}}
\newcommand{\MM}{\mat{M}}
\newcommand{\GG}{\mat{G}}

\newcommand{\jtvp}{\textbf{J\textsuperscript{T}VP}}
\newcommand{\jvp}{\textbf{JVP}}

\newcommand{\hvp}{\textbf{HVP}}

\newcommand{\sumj}{\sum_{j=1}^m}
\newcommand{\sumk}{\sum_{k=1}^K}
\newcommand{\pgrad}{\boldsymbol{\partial}}
\newcommand{\maxnorm}[1]{\left\|#1\right\|_{\text{max}}}

\newcommand{\Rel}[1]{\mathfrak{R}[#1]}
\newcommand{\Img}[1]{\mathfrak{I}[#1]}
\newcommand{\Reals}{\mathbb{R}}
\newcommand{\Complexes}{\mathbb{C}}

\newcommand{\setO}{\mathfrak{O}}
\newcommand{\setP}{\mathfrak{P}}

\newcommand{\delP}{\pgrad_{\sPP}}
\newcommand{\delO}{\pgrad_{\sOO}}

\newcommand{\grado}{\grad_{\OO}}

\newcommand{\diag}[1]{\mathtt{diag}\left(#1\right)}
\newcommand{\Dg}[1]{\mathtt{Dg}\left(#1\right)}

\newcommand{\hessl}{\grad^2 \ell}

\newcommand{\GGl}{\GG_\ell}
\newcommand{\GGg}{\GG_g}
\newcommand{\GGp}{\GG_p}

\newcommand{\JJc}{\check{\JJ}}
\newcommand{\GGlc}{\check{\GGl}}

\newcommand{\bh}{\bm{h}}
\newcommand{\bb}{\bm{b}}
\newcommand{\by}{\bm{y}}

\newcommand{\via}{\textrm{via}}
\newcommand{\eg}{\textit{e.g.},}
\newcommand{\ie}{\textit{i.e.},}
\newcommand{\apriori}{\textit{a priori}}

\newcommand{\ellp}{\ell_p}
\newcommand{\ellg}{\ell_g}

\newcommand{\oerr}{\ensuremath{\left\langle \epsilon_{_\OO} \right \rangle}}
\newcommand{\perr}{\ensuremath{\left\langle \epsilon_{_\PP} \right \rangle}}

\newcommand{\flow}{\ensuremath{\overline{n}_{\rm \textit{low}}}}
\newcommand{\fmod}{\ensuremath{\overline{n}_{\rm \textit{mod}}}}
\newcommand{\fhigh}{\ensuremath{\overline{n}_{\rm \textit{high}}}}

\begin{document}

\title{A matrix-free Levenberg-Marquardt algorithm for efficient ptychographic phase retrieval}

\author[1,*]{Saugat Kandel}
\author[2]{S. Maddali} 
\author[3]{Youssef S G Nashed}
\author[2]{Stephan O Hruszkewycz}
\author[4,5,6]{Chris Jacobsen}
\author[7]{Marc Allain}

\affil[1]{Applied Physics, Northwestern University, Evanston, Illinois 60208, USA}
\affil[2]{Materials Science Division, Argonne National Laboratory, Lemont, IL 60439, USA}
\affil[3]{SLAC National Accelerator Laboratory, Menlo Park, CA 94025, USA}
\affil[4]{Advanced Photon Source, Argonne National Laboratory, Lemont, Illinois 60439, USA}
\affil[5]{Department of Physics \& Astronomy, Northwestern University, Evanston, Illinois 
60208, USA}
\affil[6]{Chemistry of Life Processes Institute, Northwestern University, Evanston, Illinois 
60208, USA}
\affil[7]{Aix Marseille Univ, CNRS, Centrale Marseille, Institut Fresnel, Marseille, France}

\affil[*]{Corresponding author: saugat.kandel@u.northwestern.edu}
\date{}

\maketitle

\begin{abstract}
	
	The phase retrieval problem, where one aims to recover a
	complex-valued image from far-field intensity
	measurements, is a classic problem encountered in a range of imaging applications. Modern 
	phase 
	retrieval approaches usually rely on gradient descent methods in a nonlinear minimization 
	framework. 
	Calculating closed-form gradients for use in these methods is tedious work, and 
	formulating second order 
	derivatives is even more laborious. Additionally, second order techniques often require 
	the storage and 
	inversion of large matrices of partial derivatives, with memory requirements that can be 
	prohibitive for data-rich imaging modalities.
	We use a reverse-mode automatic differentiation (AD) framework to 
	implement an efficient matrix-free version of the Levenberg-Marquardt (LM) algorithm, a 
	longstanding method that finds popular use in nonlinear least-square
	minimization problems but which has seen little use in
	phase retrieval. 
	Furthermore, we extend the basic LM algorithm so that it can be applied for more 
	general constrained optimization problems (including phase retrieval problems) beyond just 
	the least-square
	applications. Since we use AD, we only need to specify the physics-based forward model 
	for a specific imaging application; the first and second-order derivative terms are 
	calculated 
	automatically through matrix-vector products, without explicitly forming the large 
	Jacobian 
	or Gauss-Newton matrices typically required for the LM method.
	We demonstrate that this algorithm can be used to solve both the unconstrained 
	ptychographic object retrieval
	problem and the constrained ``blind'' ptychographic object and probe retrieval problems, 
	under the popular Gaussian noise 
	model as well as the Poisson noise model. We compare this algorithm to state-of-the-art 
	first order 
	ptychographic reconstruction methods to demonstrate empirically that this method 
	outperforms best-in-class first-order methods: it provides  
	excellent convergence guarantees with (in many cases) a superlinear
	rate of convergence, all with a computational cost comparable to, or lower than, the 
	tested first-order algorithms.
\end{abstract}

\section{Introduction}
\label{sec:intro}

Ptychography is a coherent diffraction imaging (CDI) method based on the collection of 
diffraction intensities obtained by using a finite coherent beam to illuminate overlapping  
regions on an object \cite{hoppe_aca1_1969,hoppe_aca3_1969,hegerl_bbpc_1970}.  Following the  
development of a practical object reconstruction algorithm in 2004 
\cite{rodenburg_apl_2004,faulkner_prl_2004}, ptychography has found widespread use in imaging  
with X rays in 2D \cite{rodenburg_prl_2007} and 3D \cite{dierolf_nature_2010}, with visible 
light \cite{maiden_optlett_2010}, and electrons \cite{humphry_natcomm_2012}, and variants have 
appeared using Bragg diffraction \cite{hruszkewycz_nl_2012} and overlapping illumination 
angles instead of positions\cite{zheng_natphot_2013}. It is able to deliver images with a 
spatial resolution limited only by the scattering collected from the object (rather than the 
resolution of any optical elements used), and with both absorption and phase contrast (and 
phase contrast in Bragg-geometry CDI can in turn measure lattice strain in crystalline materials 
\cite{robinson_applss_2001}).

The ptychographic reconstruction step, wherein the object is reconstructed from the set of 
diffraction patterns, is a computational inversion step, typically iterative in approach, 
that aims to retrieve the phases of the diffracted intensities so that they are altogether 
consistent with the estimated sample. Standard phase retrieval problems are generally  
difficult computational inverse problems \cite{bates_optik_1982}, but because the   
ptychographic scan is performed with some overlap between adjacent probes, redundant 
information is extracted from each spatially localized area, which therefore provides robust  
and successful inversion (phase retrieval) for the whole sample. Nevertheless, the 
ptychographic reconstruction step remains a challenging large-scale numerical 
problem. The pioneering iterative projection methods (due to Gerchberg and Saxton 
\cite{gerchberg_optik_1972}, Fienup \cite{fienup_optlett_1978,fienup_applopt_1982}, and others 
\cite{elser_josaa_2003,luke_ip_2005, marchesini_rsi_2007}) developed in the context of 
standard CDI were not amenable to the joint phasing of the set of intensity patterns from a 
ptychography experiment. Dedicated strategies developed early on (the ptychographic iterative 
engine or PIE \cite{rodenburg_apl_2004}, and the difference map or DM approach 
\cite{thibault_ultramic_2009}) were pivotal for the early expansion of method. More recently,  
however, nonlinear optimization approaches to ptychographic reconstruction 
\cite{guizar_oe_2008} have become increasingly popular: they are more robust, accurate, and  
highly versatile as they are built explicitly on physical modeling assumptions associated with 
the experimental setup.

In ptychography, two specific situations arise that correspond to classes of inversion problems
of increasing difficulty. The first of these is the situation where an accurate estimate of
the illuminating probe is available (\eg{} \via{} a calibration performed previous to the
experiment), so that we can assume that the probe function is known. Retrieving the object from
the diffraction patterns is then equivalent to an ``unconstrained'' minimization problem.
For this case, researchers have proposed a variety of iterative solution methods, with most of
them consisting of first-order gradient-based iterations. First-order strategies
\cite{rodenburg_apl_2004,marchesini_rsi_2007,
	guizar_oe_2008,godard_oe_2012,candes_itit_2015,zhang_anips_2016, odstrcil_oe_2018,
	maiden_optica_2017} 
provide updates that are relatively easy to compute, but they are limited in their convergence 
speed \cite{qian_ipa_2014, yeh_oe_2015}. While recent second-order iterative methods 
\cite{qian_ipa_2014,yeh_oe_2015,zhang_sr_2017} address this convergence speed issue, their use 
in ptychography is strongly impaired by the high computational cost per iteration required to 
evaluate the second-order derivative matrix at each iterative update step\cite{nocedal_2006}.  
Furthermore, knowledge of the true nature of the probe as it illuminates the sample is 
difficult and in some cases impossible to ascertain with sufficient accuracy for the 
unconstrained ptychography problem. 
This leads to the common second situation where we need to retrieve (or at least refine) 
\textit{both} the object and the probe structure from the dataset: this is the ``blind'' 
ptychography problem. 
Since the set of ambiguous solutions is much larger in this second situation, the 
problem is more difficult to solve efficiently. Early algorithms developed for this purpose 
\cite{maiden_ultramic_2009,thibault_ultramic_2009} were therefore subject to stagnation
\cite{hesse_siamjis_2015,chang_sjis_2019}. More recent algorithmic developments partially solve
this issue by applying additional constrains on the probe and the sample \via{} proximal
operators \cite{hesse_siamjis_2015,chang_sjis_2019}. As we will see, while
these proximal operators are very versatile tools that have been used extensively to build
constrained minimization algorithms \cite{parikh_fto_2014}, this flexibility may be outweighed
by moderate convergence speed, so that other strategies may be preferred \cite[Chapter
3]{boyd_2011}.

The formulation of a gradient-based minimization strategy to address ptychography 
(and general optimization problems)
typically requires the manual derivation of closed-form gradient expressions. 
This is already a tedious and inflexible procedure just for a first-order minimization strategy, 
and it becomes even more difficult if we want to formulate a higher-order minimization method. An 
increasingly popular alternative to such by-hand derivations is to use the powerful 
``automatic differentiation'' or ``algorithmic differentiation'' (AD) technique for the derivative 
calculations \cite{griewank_siam_2008}. In the AD framework, once we specify the physics-based 
experimental forward model, we can 
calculate the derivatives (first or higher order) with respect to any desired model component 
automatically, without any additional mathematical manipulation. Moreover, if we modify the 
forward model, the changes are also transferred to the derivatives
automatically. This provides two key advantages for ptychography.  First, the derivatives (and associated matrix-vector 
products) related to both the object and probe variables are similarly easy to access.  
Second, we can account for a modification of the model components---such as the optical device, the propagation method, or the noise regime---with minimal effort.
Recent works have demonstrated that the AD framework can be used to conveniently and flexibly solve 
general phase retrieval problems (\eg{} ptychography, tomography, and more) through popular first-order 
minimization strategies 
\cite{jurling_josa_2014,nashed_procedia_2017,kandel_oe_2019,du_sa_2020,du_arxiv_2020}.

We propose here a generic, AD-based 
Levenberg-Marquardt (LM) minimization strategy to deal with the optimization constraints  met 
in ptychography. The proposed strategy solves both the standard and blind ptychography 
problems efficiently. The LM algorithm used here is essentially a regularized second-order 
iterative approach that offers fast convergence 
\cite{levenberg_qam_1944,nocedal_2006,hansen_jhu_2013}. 
In contrast to existing works \cite{ramos_pra_2019,ma_itit_2019}, 
	our LM implementation uses iterative updates
calculated using a computationally efficient ``matrix-free'' \cite{diamond_iccv_2015} 
fashion using the AD framework. The key to this matrix-free method is to only ever use the 
second-order derivative matrix to calculate matrix-vector
products. 
Just as the Fast Fourier Transform method calculates the Discrete Fourier Transform (DFT)
efficiently without forming the DFT matrix itself, our algorithms calculate the necessary
matrix-vector products efficiently without ever forming the full second order derivative 
matrix. We accomplish this by using a ``Hessian-free'' AD approach
\cite{pearlmutter_nc_1994,schraudolph_nc_2002,martens_phd_2016}. 

Our overall contributions in this paper are as follows:
\begin{enumerate}
    \item We modify the classical LM method so that it is based on the ``Generalized 
    Gauss-Newton'' (GGN) extension \cite{schraudolph_nc_2002} to the Gauss-Newton matrix and 
    so that it incorporates the ``projected gradient'' \cite{kanzow_jcam_2004,fan_jimo_2013} 
    extension to handle convex constraints. In contrast to the classical LM  method, which can 
    only be applied to solve unconstrained nonlinear least-squares (NLSQ) minimization 
    problems, this extended approach can be applied towards general constrained minimization 
    problems.
    \item Our LM implementation is entirely matrix-free. Since this approach is AD-based and 
    does not require closed-form derivative expressions, it can be used in a  drop-in fashion 
    within nonlinear minimization  strategies for other inverse problems.
    \item We derive analytical expressions for the diagonal elements of the  GGN 
    matrix for the ptychography application. These expressions can be used for 
    ``preconditioned'' LM iterative updates, and 
    even for preconditioned iterative updates within first-order gradient-based optimization methods (such as the nonlinear conjugate gradient method).
    These expressions are easy to adapt for other phase retrieval problems. 
    \item We demonstrate empirically that the LM method successfully solves the ptychographic 
    phase retrieval problem for both the Gaussian (NLSQ) and Poisson (non-NLSQ) 
    noise models, with a computational cost comparable to, or lower than, state-of-the-art first order methods, and in many cases (for the Gaussian noise model) even provides a superlinear rate of convergence. 
\end{enumerate}

In this paper, we first provide (\Cref{sec:problems}) a brief description of the two 
canonical problems we aim to solve, namely  the far-field ptychographic object reconstruction 
problem, and the far-field blind ptychography problem. Next, we contextualize the  
matrix-free LM strategy (\Cref{sec:LM_principles}), then detail the implementation of the 
proposed algorithm (\Cref{sec:algorithms}). Finally, we use a variety of numerical 
experiments (\Cref{sec:experiments}) to demonstrate that the LM algorithm makes for a robust 
and efficient optimization strategy to solve the ptychographic reconstruction problem.

\section{Some canonical reconstruction problems in ptychography}
\label{sec:problems}

We first provide a short description of the experimental acquisition model considerered for
the ptychographical reconstruction problems.\medskip

In the far-field 2D ptychography experiment, which is the most common variant of ptychography 
in the literature, we illuminate an unknown 2D object with a coherent probe beam localized to 
a small area on the object, and record the intensity in the far field using a pixel array 
detector. Using a raster scan of $K$ spatially overlapping illumination spots, we generate a 
sequence of $K$ diffraction patterns at the detector plane. In the following sections, we 
model the object as a 2D grid of $N_x \times N_y=N$ pixels represented by the  vector 
$\OO\in \Complexes^N$, and we model the localized probe as a grid of $M_x \times M_y=M$ pixels 
(with $M<N$) represented by the vector $\PP \in \Complexes^M$. 
At each illumination position $\rr_k$ (with $k=1,2,\dots,K$), the binary shift operator $\SSk$ 
(a $M\times N$ matrix) extracts the illuminated $M$ object pixels to generate the transmitted 
wave function $\bpsi_k\in\Complexes^M$ of
\begin{align}
\bpsi_k = \diag{\PP} \vdot \SSk \vdot \OO,
\label{eq:exit_wave}
\end{align}
where $\diag{\PP}$ is an $M\times M$ diagonal matrix containing the elements of $\PP$ in its 
main diagonal. The expected wavefield intensities at the detector plane are given by
\begin{align}
\label{eq:expected_count}
\bh_k = \abs{\hbpsi_k}^2 + \bb_k \qquad\text{with}\qquad\hbpsi_k = \FF \vdot \bpsi_k,
\end{align}
where $\FF$ is the 2D discrete Fourier transform 
operator, $\abs{\cdot}$ is the element-wise
modulus of a vector and $\bb_k$  is the (incoherent) experimental background that we shall 
assume is known and has strictly positive components\footnote{The strict positivity assumption
	is physically relevant and also ensures that the
	gradients of the error metrics (\Cref{eq:error_metric}) are 
finite everywhere. However, there also exist alternative approaches such as 
``subgradient'' formulation (see example in \cite{ma_itit_2019} for LM phase retrieval 
with the Gaussian error metric) or proximal gradient formulations (see example in  
\cite{he_arxiv_2016} for the Poisson error metric) to optimize error metrics with 
discontinuous gradients.
}. 
Since statistical fluctuations associated with the use of a finite number of illuminating 
photons are inherent to the measurement process,  the recorded data $\by_k$ necessarily differ 
from the expected values of \Cref{eq:expected_count}. In the non-linear minimization 
approach, which we apply in this work, we account for the noise by defining an 
``error metric''  (or fitting function). The generic form of the error metric is 
\begin{align}
f = \sumk \sum_{m=1}^M \LL(\,\cdot \,,\, y_{k,m}) \circ h_{k,m}
\label{eq:error_metric}
\end{align}
with $y_{k,m}\in \yy_k$, $h_{k,m}\in \bh_k$, and where 
$\LL : \mathbb{R} \times \mathbb{R} \rightarrow \mathbb{R}$ 
is derived from the specific model chosen for the noise-driven fluctuations in the 
measurements. We usually require that $\LL$ is a strictly convex functional, thereby defining 
a proper ``metric''\footnote{Because the functionals derived from the maximum-likelihood principle do not define strict 
distances in general (\eg{} the Poisson error metric is not symmetric and can have negative 
values), the term ``metric'' should be understood in a somewhat loose sense.}
between the actual (noisy) measurement $y_{k,m}$ and the expected quantity $h_{k,m}$ (holding 
in average only).
The additive structure of  \Cref{eq:error_metric} allows us to account for all the 
measurements and produces a single real value which acts as a figure of merit. Concerning the 
noise model, the Poisson distribution  is the natural choice for a photon counting process, so 
we adopt it here. Following the maximum likelihood principle introduced by Fisher 
\cite{fisher_1956}, we obtain the error metric 
\cite{thibault_njp_2012,godard_oe_2012,odstrcil_oe_2018,chang_sjis_2019}
\begin{align}
\ffp := \sumk\sum_{m=1}^M h_{k,m} - y_{k,m} \log h_{k,m}.
\label{eq:poisson_nll}
\end{align}
In the low photon-count regime, the above error metric is often the metric of choice because 
it reduces the estimation biases (systematic errors) in CDI reconstructions 
\cite{godard_oe_2012}. As the expected count $h_{k,m}$ gets higher, we can use the frequently-used
least-square functional as a consistent 
approximation of \Cref{eq:poisson_nll} to give
\begin{align}
\ffg := \frac{1}{2}\sumk \sum_{m=1}^M  \left( h_{k,m}^{1/2} -  y_{k,m}^{1/2} \right)^2
\,  = \, \frac{1}{2} \sumk \left\Vert \, \bh_k ^{1/2} - \yy_{k}^{1/2} \,\right\Vert^2
\label{eq:magnitude_nls}
\end{align}
where $||\cdot||$ is the usual Euclidean norm and ${\cdot}^{1/2}$ is the element-wise square 
root. This latter metric is derived from an additive perturbation model over the recorded 
magnitudes $y_{k,m}^{1/2}$ with the assumption that the perturbation follows a Gaussian 
distribution with a constant variance. For low to moderate photon-count regimes, we expect a 
greater bias in the reconstructions obtained using \Cref{eq:magnitude_nls} than in those using 
\Cref{eq:poisson_nll}. However, we shall see in \Cref{sec:experiments} that the standard 
least-square metric of \Cref{eq:magnitude_nls} leads to a faster convergence in general, and 
essentially identical results in the high-count regime. 

In the non-linear minimization paradigm, our solution of the ptychographical reconstruction 
problem is implicitly defined \via{} the minimization of either \Cref{eq:poisson_nll} or 
\Cref{eq:magnitude_nls} with respect to the unknown quantities (and possibly under 
additional constrains). More specifically, if the structure of the probe beam is known, the 
object can be retrieved by numerically solving the unconstrained minimization problem 
\begin{align}
\OO_\star \in \argmin_{\OO\in\Complexes^N} f_{\bullet}(\OO)
\label{eq:cdp}
\end{align}
where $f_{\bullet}$ stands for $\ffp$ or $\ffg$, and where the dependence on the quantity of 
interest $\OO$ was given by \Cref{eq:expected_count,eq:exit_wave}. For the sake of 
clarity, we call \Cref{eq:cdp} the ``standard ptychographic reconstruction'' problem 
(SPR). 
The problem above, equipped with the functional $\ffg$, is an example of the ``coded 
diffraction pattern'' problem that has been extensively analyzed in recent phase retrieval  
literature \cite{candes_itit_2015,zhang_anips_2016,qian_ipa_2014}. 

If the structure of the probing field is not completely known, we need to 
retrieve the object as well as the probe from the diffraction dataset. This ``blind 
ptychographic reconstruction'' problem (BPR) is structurally distinct from 
the SPR problem (or the general phase retrieval problem): it allows for a much larger set of ambiguous solutions, and is thus much more difficult than the SPR problem \cite{hesse_siamjis_2015,fannjiang_ip_2019}.
Therefore, the recent literature solves the BPR problem by taking into account not only the 
diffraction dataset but also \apriori{} information about the object and the probe, such 
as spatial or spectral support, non-negativity, or magnitude
constraints, typically via proximal operators \cite{hesse_siamjis_2015, chang_sjis_2019, fannjiang_ip_2019}. In this approach, the BPR problem is then a constrained 
minimization problem of
\begin{align}
(\OO_\star, \PP_\star) \in \argmin_{\OO\in\setO,\PP\in\setP} f_\bullet(\OO,\PP)
\label{eq:bp}
\end{align}
where $\setO\subset\Complexes^N$ and $\setP\subset\Complexes^M$ are 
closed convex sets associated with the object or probe constraints respectively. In
keeping with the existing blind ptychography literature, we only report numerical
results for our constrained formulation of the BPR problem. If these constraints are not applied, 
the possible solution space becomes very large (due to scaling ambiguities) and the BPR problem 
can be harder to solve.

We can use either the Poisson error metric of \Cref{eq:poisson_nll} or the Gaussian
error metric of \Cref{eq:magnitude_nls} to solve either the SPR case (\Cref{eq:cdp})  or 
the BPR case (\Cref{eq:bp}). These problem instances thus define the four canonical problems 
that we aim to solve \via{} a fast and computationally efficient algorithm.

\section{The principle of a matrix-free Levenberg-Marquardt strategy}   
\label{sec:LM_principles}

As in any phase retrieval problem, both the SPR and  BPR problems are NP-hard and, in general, 
cannot be solved exactly in polynomial time \cite{wang_um_2018}. 
We thus resort to ``gradient-based'' minimization strategies with good 
local convergence properties, \ie{} algorithms that ensure that any 
stationary point is a local minimizer of the considered problem.
Since the considered error metrics\footnote{In this section, for the sake of notational simplicity, $f_\bullet$ (standing for $\ffp$ or 
$\ffg$) is simply denoted $f$. 
Furthermore, the generic complex-valued vector $\zz \in \mathbb{C}^n$ is used to denote
the set of parameters we want to retrieve: we then have $\zz\equiv\OO$  with  $n=N$ for the 
SPR problem and $\zz \equiv(\OO,\PP)$ with $n=N+M$ for the BPR problem.
We also introduce the notation $m=KM$ as a shorthand to denote the total number of available 
measurements after the ptychographic scan.
}
 $f$ are defined over a set of complex-valued parameters $\zz\in \mathbb{C}^n$, 
we rely on the Wirtinger (or $\Complexes\Reals$ calculus) extension to the notion 
of the derivative, wherein we regard $f$ as a function of \textit{two} variables
\cite{brandwood_iproc_1983,kreutz_arxiv_2009,sorber_siam_2012}. 
We accomplish this by writing $f$ as a function of the real and 
imaginary parts of $\zz$, where $\Rel{\cdot}$ and $\Img{\cdot}$ denote the element-wise 
operations\footnote{An alternative, formally equivalent, approach is to write $f$ as a 
function of $\zz$ and its element-wise complex conjugate $\zz^*$.}.
To clarify that the optimization is accomplished entirely \via{} real-valued coordinates, we 
define the new vector $\zzt \equiv [\Rel{\zz}^T, \Img{\zz}^T]^T\in \mathbb{R}^{2n}$
so that $f(\zzt) \equiv f(\zz)$. The gradient $\grad f(\zzt)$ and any subsequent higher-order 
derivatives are then all real-valued. With this definition in hand, we can now derive any 
gradient-based nonlinear minimization method from a second order expansion of $f$ around an 
arbitrary point $\zzt$ \cite{nocedal_2006}: 
\begin{align}
f (\zzt + \Delta \zzt) \approx  
f(\zzt) + \grad f(\zzt)^T \vdot \Delta \zzt + \frac{1}{2}\Delta \zzt^T \vdot \MM(\zzt) \vdot \Delta \zzt
\label{eq:taylor_expansion}
\end{align}  
where $\MM(\zzt)$ is a matrix of size $2n\times 2n$ that describes the
local ``curvature'' of the objective function.
While the canonical quadratic approximation of $f(\zzt)$ uses $\MM(\zzt) = \grad^2 f(\zzt)$, 
with $\grad^2 f(\zzt)$ the ``Hessian'' matrix, we can also use other choices of the 
matrix $\MM(\zzt)$  to get alternative quadratic approximations of $f(\zzt)$. As long as our 
choice of the curvature matrix $\MM(\zzt)$ is positive semi-definite, the quadratic 
approximation obtained is easy to minimize, and the minimizing step thus obtained can be used 
to define a descent step for $f(\zzt)$. In fact, different choices for  $\MM$ give different 
descent steps, and, as such, comprise different optimization methods, as we discuss in the 
following sections.

\subsection{From first-order to second-order methods}
\label{subsec:theory_first_order}

We can make a simple choice for the quadratic approximation in \Cref{eq:taylor_expansion} 
by discarding any anisotropy and coupling in the local curvature of $f$ and setting 
$\MM=\tfrac{1}{\alpha}\II$, where $\alpha>0$ controls the curvature magnitude. This choice 
leads to the update step
\begin{align}
	\Delta \zzt_\star = -\alpha \grad f(\zzt)
\label{eq:steepest_descent}
\end{align}
which is none other than the ``steepest descent'' update step. 
In the phase retrieval context, a number of classic algorithms (such as ER 
\cite{fienup_applopt_1982}), as well as some recent algorithms (such as Wirtinger flow 
\cite{candes_itit_2015}, reshaped Wirtinger flow \cite{zhang_anips_2016}, and others) can 
be interpreted as variations of this steepest descent method, just with different
initializations and error metrics. There also exist stochastic minibatch (or 
sub-sampled) variations of these steepest descent methods that use a subset of the full 
dataset to calculate each update (such as PIE/ePIE \cite{rodenburg_aiep_2008} and 
minibatch reshaped Wirtinger flow\cite{zhang_anips_2016}). These methods are easy to implement 
but can require a large number of iterations (the number of which depends strongly 
on the step size $\alpha$) to converge to a solution. Even if we choose the optimal step size 
at every update, these algorithms exhibit, at best, a linear rate of convergence, unlike 
second-order methods which can provide superlinear or quadratic rates of convergence 
\cite{nocedal_2006,bertsekas_athena_1999}.

In the optimization literature, the simple steepest descent scheme of 
\Cref{eq:steepest_descent} has, to a large extent, been superseded by more sophisticated 
``accelerated'' first-order optimization techniques. Many of these methods have also been 
applied to the phase retrieval problem: these include nonlinear conjugate gradient methods 
\cite{fienup_applopt_1982,marchesini_rsi_2007,thibault_njp_2012,wei_josaa_2017}, heavy-ball 
momentum and Nesterov's accelerated gradient (NAG) methods 
\cite{maiden_optica_2017,pauwels_itsp_2018,xu_arxiv_2018}, and the Adam method 
\cite{ghosh_iccp_2018,kandel_oe_2019}. While these algorithms are easy to implement and have 
low per-iteration computational cost, they are essentially attempting to adapt to the geometry of 
$f(\zzt)$  by utilizing only the gradient information (from current and prior iterations) 
\cite{kingma_corr_2014}.
Consequently, to the degree that such approximations do not accurately capture the local curvature, 
such algorithms will display less improvement per iteration than a pure second-order 
optimization method.

To develop a second-order optimization method, we rely on the canonical quadratic 
approximation of \Cref{eq:taylor_expansion}, with $\MM(\zzt) = \grad^2 f(\zzt)$ as the 
Hessian matrix, to fully capture the local curvature information at $\zzt$. Assuming that 
this Hessian is full rank, the so-called Newton's step minimizing the resulting quadratic 
approximation is $\Delta\zzt_\star = - \minv{[\grad^2 f(\zzt)]} \vdot \grad f(\zzt)$.
Provided $\zzt$ is close enough to a local isolated minimizer, the Newton's approach utilizes 
the full local curvature captured in $\grad^2 f(\zzt)$ to attain a fast (typically quadratic)
rate of convergence \cite[Chapter 3]{nocedal_2006}. This very good local convergence property 
is generally offset by prominent, long-known, robustness issues \cite[Chapter 6]{nocedal_2006}. 
In particular, when $\grad^2 f(\zzt)$ is not positive semi-definite (PSD), which is often the 
case for non-convex objective functions, the step may not be a descent step at all. To address 
this issue, we can resort to a PSD approximation of the Hessian, which is the idea behind the 
``Generalized Gauss-Newton'' method.

\subsection{From Gauss-Newton to constrained Levenberg-Marquardt}
\label{subsec:theory_lm}

The classical Gauss-Newton matrix arises in the following form in the context of nonlinear 
least-squares minimization problems, and can be derived when the error metric takes the form 
$\LL(\hh) = \tfrac{1}{2}\norm{\hh - \yy}^2$ so that the objective function $f$ reads as
\begin{align}
f(\zzt) &= \frac{1}{2}\norm{\hh(\zzt) - \yy}^2 
= \frac{1}{2} \sumj \left(h_j(\zzt) - y_j\right)^2.
\label{eq:lsq}
\end{align}
Let us introduce the ``Jacobian matrix'' $\JJ(\zzt)\in\Reals^{m\times 2n}$ defined 
element-wise as $[\JJ(\zzt)]_{ij} = \pdv*{h_i}{\zzt_j}$. The first- and second-order 
derivatives of \Cref{eq:lsq} can be shown \cite{nocedal_2006,hansen_jhu_2013} to be
\begin{equation}
\left\{
\begin{split}
\grad f(\zzt) & \,= \, \JJ^T (\zzt) \vdot \left(\hh - \yy\right) \,=\,  \JJ^T (\zzt) \vdot \grad_{\hh} \LL(\hh),\\
\grad^2 f(\zzt) & \,=\, \JJ^T (\zzt)\vdot \JJ(\zzt) + 
{\textstyle \sumj} (h_j - y_j) \grad^2 h_j,
\end{split}
\right.
\label{eq:lsq_hessian}
\end{equation}
where $\grad_{\hh} \LL$ is the gradient of $\LL$ with respect to $\hh$, and $\grad^2 h_j$ is 
the Hessian of the $j$-th component of $\hh$ with respect to $\zzt$. When the 
``residuals'' $(h_j - y_j)$ are small, or when the model $\hh$ is almost linear locally 
so that $\grad^2 h_j$ terms are small, the PSD Gauss-Newton (GN) matrix $\GG(\zzt) 
:=\JJ(\zzt)^T \vdot \JJ(\zzt)$ is a close approximation of the local Hessian. In a GN 
optimization algorithm, the quadratic surrogate \Cref{eq:taylor_expansion} is built 
with $\MM(\zzt) \equiv \GG(\zzt)$ so that the update $\Delta \zzt_\star$ 
is the solution of the linear system
\begin{align}
	\label{eq:gn_update}
	\GG(\zzt) \vdot \Delta \zzt_\star = -\grad f(\zzt) = -\JJ^T (\zzt) \vdot \grad_{\hh} \LL(\hh).
\end{align} 

In the early 2000's, Schraudolph \cite{schraudolph_nc_2002} generalized the GN method to 
arbitrary error metrics $\LL$ that are convex in $\hh$. For any such $\LL$, the Generalized 
Gauss-Newton (GGN) matrix is defined as
\begin{align}
\GG(\zzt)=\JJ(\zzt)^T \vdot \grad_{\hh}^2 \LL (\hh) \vdot \JJ(\zzt)
\label{eq:ggn}
\end{align}
where $\grad_{\hh}^2 \LL$ is the Hessian of $\LL(\hh)$ with respect to $\hh$. The Hessian of 
$f(\zzt)$ then reads 
\begin{align}
\grad^2 f(\zzt) = \GG(\zzt) + 
\sumj [\grad_{\hh} \LL(\hh)]_j  \grad^2 h_j
\label{eq:ggn_approximation}
\end{align}
where $\GG(\zzt)$ is now given by \Cref{eq:ggn}.  Similar to the classical GN derivation, the 
GGN strategy drops the second term in the  above Hessian to build the update. The GGN step 
$\Delta\zzt_\star$ minimizing the surrogate \Cref{eq:taylor_expansion}  solves the linear 
equation $\GG(\zzt) \vdot \Delta \zzt_\star = -\grad f(\zzt)$. Since $\LL(\hh)$ is convex, the 
GGN matrix $\GG(\zzt)$ is PSD and the solution $\minv{-\GG(\zzt)} \vdot \grad f(\zzt)$ 
provides a well-behaved, locally decreasing update direction. This basic GGN method provides a 
number of advantages over first-order methods as well as Newton's method in many 
situations. However, if the residuals of \Cref{eq:lsq} are large, or if the Jacobian matrix is 
ill-conditioned  \cite{nocedal_2006,hansen_jhu_2013}, the method may be unstable.  To address 
this deficiency, we explore a popular variation of the basic GN algorithm: the  
Levenberg-Marquadt method.

In the classical Levenberg-Marquardt (LM) algorithm, we obtain the iterative updates by 
solving the linear equation
\begin{align}
	\left(\GG(\zzt) + \lambda \II\right)\vdot \Delta \zzt_\star = -\JJ(\zzt)^T \vdot \grad_{\hh}\LL(\hh),
	\label{eq:lma}
\end{align}
which is an interpolation between the steepest descent update (\Cref{eq:steepest_descent})
and the GN update (\Cref{eq:gn_update}), with $\lambda > 0$ the interpolation parameter 
\cite{levenberg_qam_1944,marquardt_jsiam_1963,more_na_1978,nocedal_2006}. 
The value of $\lambda$, which we adjust at every step, indicates the extent to which we trust 
the minimizer of the GN quadratic approximation to minimize the true objective function 
$f(\zzt)$. When $\lambda$ is very small, the GN term dominates, and the LM step is 
approximately along the GN update direction; when $\lambda$ is very large, the LM step is 
approximately along the steepest descent direction (with the step size $1/\lambda$). This 
adjustment allows the algorithm to identify minimizing steps even if the residuals are large 
or if the Jacobian matrix is ill-conditioned. 

The LM algorithm has been established as a workhorse for nonlinear
least-squares minimization applications (with the classical GN matrix). Recent works have 
extended the LM method by including a ``projected gradient'' \cite{bertsekas_athena_1999} 
approach within the LM framework \cite{kanzow_jcam_2004,fan_jimo_2013} to ensure convergent 
descent for minimization problems with convex constraints. Furthermore, the LM algorithm has 
also been successfully applied to minimization problems with more general error metrics (with 
the GGN matrix) \cite{laurence_nm_2009,huang_jcm_2017}. 

For this work, we implement a generalized LM algorithm that includes the GGN adaptation along 
with the projected gradient extension; this enables its use in general minimization problems 
with convex constraints.  We present the algorithmic details in \Cref{sec:algorithms}. 
 
\subsection{A truncated, matrix-free Levenberg-Marquardt method}
\label{subsec:TMFLM}

At first glance, the LM method inherits computational bottlenecks that have long been 
associated with second-order optimization strategies: the computational and memory costs 
required to calculate the curvature matrix grow quadratically with the problem 
dimension. Indeed, at each LM iteration we need to calculate the full $2n\times 2n$ element 
GGN matrix of \Cref{eq:lma}, which can be prohibitively expensive even for moderate problem 
dimensions. In this work, we circumvent this computational difficulty by using two key 
ingredients that work in conjunction.

A first thrust toward a computationally efficient LM approach is to resort to a ``truncated'' 
version of the method, which means that we give up on the idea of calculating the LM step 
$\Delta \zzt_\star$ by exactly solving the linear relation in \Cref{eq:lma} 
\cite{wright_anziam_1985, dan_oms_2002}. Instead,  we compute an \textit{inexact} (but 
sufficiently accurate in practice) solution by using an iterative solver. To achieve this, we 
use the conjugate gradient (CG) method, which has been successfully used since the 1980's in 
very similar situations (see \cite{nash_joam_2000} and references therein). The resulting LM 
method now contains a minor (nested) loop solving for $\Delta \zzt_\star$ in each LM update. 
This nested loop is stopped when the following stopping rule is met:
\begin{align}
	\norm{\left(\GG(\zzt) + \lambda \II\right)\vdot \Delta \zzt + \JJ(\zzt)^T \vdot \grad_{\hh}\LL(\hh)} \leq \eta \norm{\JJ(\zzt)^T \vdot \grad_{\hh}\LL(\hh)},
	\label{eq:lma_termination}
\end{align}
where  $\eta \in (0, 1)$. This termination condition ensures that the LM procedure remains 
globally convergent with, under optimal conditions, a superlinear local convergence rate
\cite{huang_jorsc_2018}. 

The nested solver does not prevent the memory requirement from being prohibitive by itself. 
For instance, each CG iteration requires a matrix-vector multiplication that involves the full 
GGN matrix. For ``real-world'' ptychographic problems, the storage of such a matrix is not a 
realistic option. We solve this second computational bottleneck by using the AD
framework, as it allows the method to be matrix-free in that
none of the GGN matrices involved in the LM iterations are effectively stored or even built. 
Instead, for some $\bm{v}\in\Reals^{2n}$, we directly access the 
``matrix-vector-products'' $\GG(\zzt)\vdot \bm{v}$ required in the linear (CG) solver 
\via{} the reverse-mode AD method. Calculating these matrix-vector products through 
such a matrix-free method requires a computational cost that is larger than that for 
the gradient calculation by only a small multiplicative factor \cite{townsend_blog_2017}. 
\Cref{appendix:AD_mvp} details the mechanism at work with reverse-mode AD  to compute 
matrix-free, generic vector multiplication with Jacobian and Hessian operators.

A few additional considerations are required to deal with poorly scaled 
problems, \ie{} problems where the changes to $\zzt$ in a certain direction produce much 
larger variations in the value of $f$ than do changes in other directions of $\zzt$ 
\cite[Chapter 2]{nocedal_2006}.  In such a situation, the GGN matrix in \Cref{eq:lma} is 
ill-conditioned and the CG solver may not find a viable update direction even after a large 
number of iterations.  
The resulting LM updates may converge slowly, or even fail to converge. One way to
address this scenario is by replacing the identity matrix in \Cref{eq:lma} by the diagonal 
matrix $\bm{D} = \Dg{\GG(\zzt)}$, where $\Dg{\cdot}$ is a diagonal matrix built from the main 
diagonal of the square matrix given as an argument. This modification makes the algorithm 
invariant under diagonal scaling of the variables \cite{more_na_1978}.  In addition, we may 
also use the preconditioned CG (PCG) method to substantially accelerate the convergence of the
linear solver computing the inexact update $\Delta \zzt_\star$. In this work, we follow an 
existing example \cite{martens_phd_2016} and test the simple diagonal (Jacobi) preconditioner
$\Dg{\GG(\zzt) + \lambda \bm{D}}$ (calculated analytically in \Cref{appendix:jacobi_lm}), and 
find that it provides efficient convergence.

\section{Matrix-free LM algorithm for the canonical problems}
\label{sec:algorithms}

We now provide a detailed description of the matrix-free truncated LM
algorithms used in this work. For notational simplicity, as in the previous
section, the vector $\zzt$ is used hereafter as a generic short-hand 
for the set of $\mathbb{CR}$-valued parameters that are optimized,
\ie{} we have $\zzt = \tilde{\OO}$ for the SPR or $\zzt = (\tilde{\OO},\tilde{\PP})^T$
for the BPR problem.

\subsection{From intensity to magnitude-based LM updates}
\label{subsec:ampl_error_metrics}

The LM method has long been established as a versatile, fast and provably 
convergent solver for NLSQ minimization \cite{nocedal_2006,hansen_jhu_2013}. 
In addition, as explained in \Cref{subsec:theory_lm}, the method can be extended beyond the 
usual quadratic error metrics \via{} the GGN formulation. As a direct consequence, both the 
error metrics $\ffg$ and $\ffp$ defined in \Cref{sec:problems} can be minimized within this 
framework. 
We also gather from \Cref{subsec:theory_lm} that, for a given error-metric $f_\bullet$, 
the LM update is not unique: any functional decomposition of the error-metric
that preserves a positive definite central part in the GNN matrix defines a legitimate, yet 
specific LM strategy minimizing $f_\bullet$ \cite{schraudolph_nc_2002}.
For instance, we define \Cref{eq:error_metric} \via{} the intermediate (physically relevant) 
intensity variables  $h_{k,m}$, hence providing ``intensity-based'' formulations of the 
error-metrics \Cref{eq:poisson_nll,eq:magnitude_nls}. Since $\LL$ is strictly 
convex, the functional decomposition \Cref{eq:error_metric} suggests a straightforward,
intensity-based LM update. However, as it is often the case in mathematical programming, 
numerical implementations derived from equivalent mathematical formulations can differ 
substantially in performance in solving the very same problem.  

We see this difference in action when we introduce the magnitude of the expected
diffracted wave-field\footnote{This is a slight abuse 
	of notation: since the background noise is incoherent, 
	$h_j^{1/2} = \left(\abs{\FF \vdot \hbpsi_k}^2 + \bb_k\right)^{1/2}$ 
	is not strictly the magnitude of a complex wavefront.
}, defined as
\begin{align}
	\label{eq:magnitude}
	\zeta_{j} := h_{j}^{1/2},
\end{align}
where $j\in \{1\cdots K\times M\}$ is a single index spanning 
both the probe position index and the pixel index. We can now introduce equivalent 
magnitude-based formulations of the intensity-based error metrics in 
\Cref{eq:poisson_nll,eq:magnitude_nls}
\begin{align}
	\overbrace{f_\bullet^\LL =  
		\sumk \sum_{m=1}^M \LL(\,\cdot \,,\, y_{k,m}) \circ h_{k,m}}^\text{intensity-based} 
	\,=\, 
	\overbrace{\sumk \sum_{m=1}^M \ell(\,\cdot \,,\, y_{k,m}) \circ \zeta_{k,m}
	= f_\bullet^\ell}^\text{magnitude-based}
	\label{eq:error_metric_magnitude}
\end{align}
where
\begin{align}
	\label{eq:ell_magnitude}
	\ell(\zeta ,\, y) := 
	\left[
	\begin{array}{ll}
		\ellp(\zeta ,\, y) =\zeta^2 - y \log \zeta^2 & \text{(Poisson)}\\
		\ellg(\zeta ,\, y) =  \frac{1}{2}\left(\zeta - y^{1/2} \right)^2 & \text{(Gaussian)}\\
	\end{array}.
	\right.
\end{align}
We note that $f_\bullet^\LL(\zzt)$ and $f_\bullet^\ell(\zzt)$ are just different functional 
decompositions of the exact same algebraic expression. As such, they also have identical 
gradient values at all points ($\grad f_\bullet^\LL=\grad f_\bullet^\ell$). However, 
we can now define two separate matrices
\begin{align}
\GG^\LL(\zzt) &= \JJ^T_h (\zzt) \vdot \grad^2_{\hh} \LL(\hh) \vdot \JJ_h(\zzt) & \text{(intensity-based)}
\label{eq:ggn_intensity}\\
\GG^{\ell}(\zzt) &= \JJ_{\zeta}^T (\zzt) \vdot \grad^2_{\bzeta} \ell(\bzeta) \vdot 
\JJ_{\zeta}(\zzt) & \text{(magnitude-based)} \label{eq:ggn_magnitude} 
\end{align}
where $\JJ_h$ and $\JJ_\zeta$ are the Jacobian matrices defined as $\left[\JJ_h 
(\zzt)\right]_{ij} = \pdv*{h_i}{\zzt_j}$ and $\left[\JJ_\zeta (\zzt)\right]_{ij} = 
\pdv*{\zeta_i}{\zzt_j}$ respectively, and $\grad^2_{\hh} \LL(\hh)$ and $\grad^2_{\bzeta} 
\ell(\bzeta)$ are the Hessians for $\LL$ and $\bzeta$ respectively. Since $\LL$ and $\ell$ are 
both strictly convex functions, both the matrices $\GG^\LL(\zzt)$ and $\GG^{\ell}(\zzt)$ can 
be interpreted as GGN matrices, and can therefore be used to formulate two different LM 
minimization strategies. 

\begin{figure}[th]
	\centering
	\includegraphics[width=0.5\linewidth]{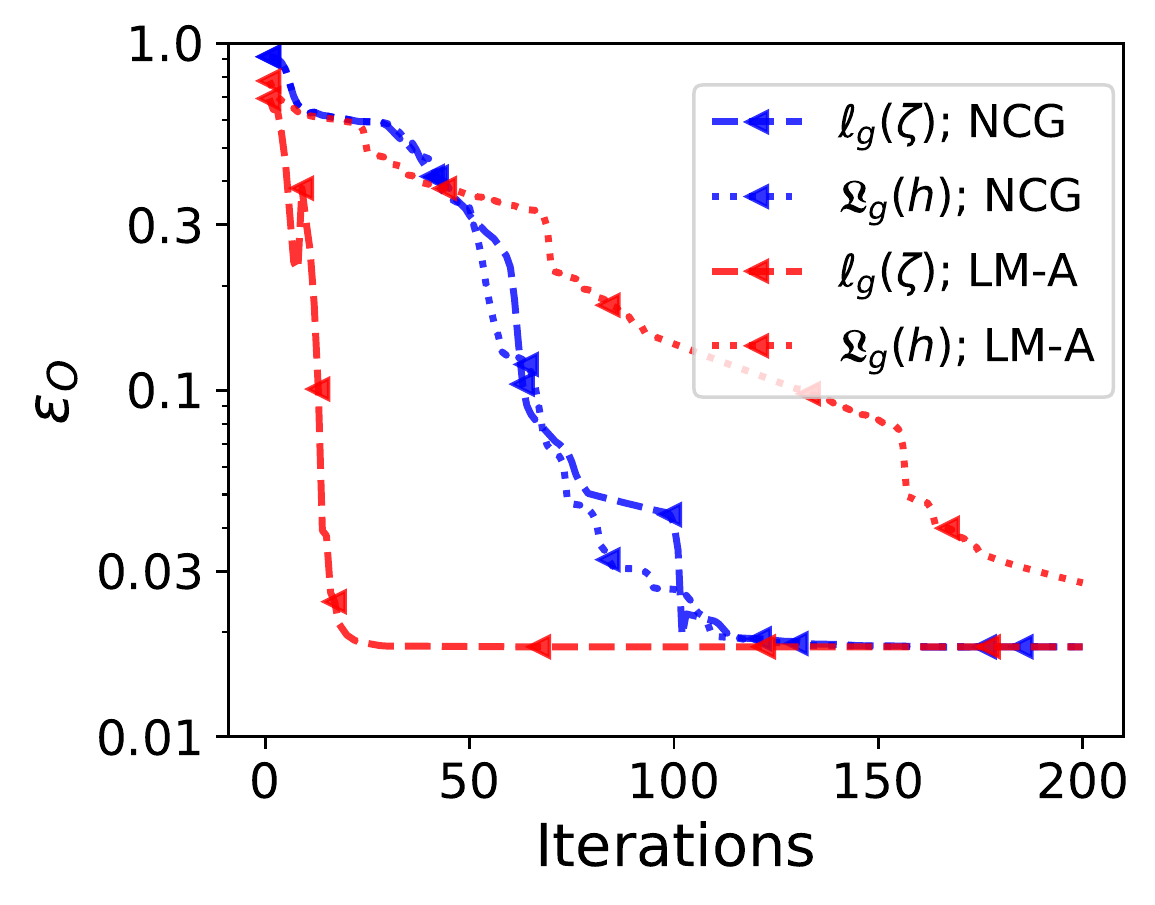}
	\caption{Comparison of the object reconstruction error for optimization with the intensity-based ($\LLg$) and magnitude-based ($\ellg$) error metric formulations (\Cref{eq:error_metric_magnitude}) for the standard ptychographic object reconstruction (SPR) problem  with a known probe. The numerical experiment uses  the parameters described in \Cref{sec:experiments}  
	for the \fhigh{} setting. For the LM algorithm, $\ellg$ enables much faster convergence than the $\LLg$, while the convergence rate of the NCG method is the same for both the metrics.}
	\label{fig:loss_magnitude_intensity}
\end{figure}
When we closely examine the expressions for $\GG^\LL(\zzt)$ and $\GG^{\ell}(\zzt)$,  
we find that $\GG^{\ell}(\zzt)$ more accurately approximates the true Hessian $\grad^2 
f(\zzt)$ than $\GG^\LL(\zzt)$ does. 
In practice, this means that the magnitude-based LM strategy shows a much faster rate of 
convergence than 
the intensity-based strategy,as we can see in the numerical results shown in \Cref{fig:loss_magnitude_intensity}; we analyze this result in 
\Cref{appendix:LM_magnitude_intensity}. For convenience, in the following sections,
we strictly use only the magnitude-based LM strategy and use the simplifying notations 
$\JJ(\zzt)\equiv 
\JJ_\zeta (\zzt)$ and $\GG(\zzt)\equiv\GG^{\ell}(\zzt)$. As a result, the linear system to be 
solved is
\begin{align}
	\left(\GG(\zzt) + \lambda \bm{D}\right)\vdot \Delta \zzt_\star
	= -\JJ(\zzt)^T \vdot \grad_{\bzeta} \ell(\bzeta)
	\label{eq:lma_magnitude}
\end{align}

Thanks to the AD implementation, all of the various linear systems derived from
our magnitude-based formulation of \Cref{eq:lma_magnitude}
can be solved iteratively in a matrix-free fashion and none of the
derivatives involved have to be obtained analytically beforehand.  
We can also use the diagonal matrix $\bm{D} = \Dg{\GG(\zzt)}$ 
in the left-hand side of the LM system of 
\Cref{eq:lma_magnitude} to deal specifically with poorly-scaled 
problems, as shown in Sec.~\ref{subsec:TMFLM}.

\subsection{Implementation of the truncated, matrix-free Levenberg Marquardt approach}
We provide below, in \Cref{alg:lm_inexact},  the main solver we use to address both the cases 
of the SPR (\Cref{eq:cdp}) and BPR (\Cref{eq:bp}) canonical problems that we introduced in 
Sec.~\ref{sec:problems}.  
\begin{algorithm}[H]
	\begin{algorithmic}[1]
		\Require $\zzt_0$, $\mu_0 > \mu_{min} > 0$, $\kappa > 1$, $\nu \in [1, 2]$, $0 < \rho_{min}\ll 1$.
		\State Set $\rho_0 = 0$.
		\For{$t=0$ to $T$}
		\State Calculate $\GG(\zzt_t)$, $\JJ(\zzt_t)$, $\grad{\ell(\bzeta_t)}$, and $\bm{D}_t$.
		\While{$\rho_t < \rho_{min}$}
		\If{$\bm{D}_t=\II$} 
		\State Set $\lambda_t = \mu_t \norm{\grad{\ell(\zeta_t)}}^\nu$ 
		\Else 
		\State  Set $\lambda_t = \mu_t$
		\EndIf
		\State Calculate the PCG preconditioner $\bm{P}_t$.
		\State Solve for $\Delta \zzt_t$ using
		\Cref{eq:lma_magnitude,eq:lma_termination_magnitude}. \label{ligne12}
		\State Calculate actual and predicted reductions
		\abovedisplayskip=0pt
		\belowdisplayskip=0pt
		\begin{align*}
			\Delta f_t^a &= f(\zzt_t) - f(\zzt_t + \Delta \zzt_t)\\
			\Delta f_t^p &= \frac{1}{2}\Delta \zzt_t^T \vdot (\JJ(\zzt_t)^T \vdot \grad \ell(\zzt_t - \lambda_t \bm{D}_t\vdot \zzt_t)
		\end{align*}
		\State Set $\rho_t = \Delta f_t^a / \Delta f_t^p$.
		\If{$\rho_t > 0.75$}
		\State $\mu_{t+1} = \max(\mu_t / \kappa,\, \mu_{min})$ and $\zzt_{t+1} = \zzt_t + \Delta \zzt_t$.
		\ElsIf{$\rho_t > 0.25$}
		\State $\mu_{t+1} = \mu_t$ and $\zzt_{t+1} = \zzt_t + \Delta \zzt_t$.
		\ElsIf{$\rho_t > \rho_{min}$}
		\State $\mu_{t+1} = \kappa \mu_t$ and $\zzt_{t+1} = \zzt_t + \Delta \zzt_t$.
		\EndIf
		\EndWhile
		\EndFor
	\end{algorithmic}
	\caption{Truncated LM algorithm: common structure for both the 
		SPR and the BPR problems. For the BPR problem, the
		complementary unit given in \Cref{alg:projected_LM} ensures 
		the constrained update.}
	\label{alg:lm_inexact}
\end{algorithm}
This LM algorithm implements the magnitude-based formulation given in the preceding section
by combining the basic truncation algorithm from \cite{huang_jorsc_2018} with the diagonal 
scaling from \cite{more_na_1978}, the GGN extension from \cite{huang_jcm_2017}, 
and the projected gradient plug-in from \cite{fan_jimo_2013}. 
As explained in Sec.~\ref{subsec:TMFLM}, each LM update rests on the inexact, matrix-free solving of the linear system 
of \Cref{eq:lma_magnitude} using  PCG with the diagonal preconditioner 
$\bm{P} = \Dg{\GG(\zzt) + \lambda  \bm{D}}$. We use a warm start procedure 
to initialize the PCG inner loop, and control the accuracy of the inexact solution \via{}
the termination condition
\begin{align}
	\norm{\left(\GG(\zzt) + \lambda \bm{D} \right)\vdot \Delta \zzt + \JJ(\zzt)^T \vdot \grad_{\bzeta}\ell(\bzeta)} \leq \eta \norm{\JJ(\zzt)^T \vdot \grad_{\bzeta}\ell(\bzeta)}
	\label{eq:lma_termination_magnitude}
\end{align}
with $\eta$ calculated according to \cite{fan_jimo_2013},
\begin{align}
	\eta = \min \left(\beta, \sqrt{\norm{\JJ(\zzt) \vdot \grad_{\bzeta}\ell(\bzeta)}} \right),
	\label{eq:lma_termination_eta}
\end{align}
where the parameter $\beta>0$ determines the accuracy of the PCG update and thereby also the 
balance between the number of inner CG iterations and outer LM iterations. Individual (inner) 
PCG iterations are much cheaper than the outer LM iterations, so it makes sense to prioritize the
accurate solution of \Cref{eq:lma_termination_eta}.  After some testing, we set 
$\beta=0.1$ to provide a good balance between the inner/outer iterations for the Gaussian
error metric. However, since the GGN matrix for the Poisson error metric is less accurate than 
that for the other error metrics, and since the LM linear system for this metric may be 
difficult to solve accurately, emphasizing highly accurate PCG updates in this case is 
likely to impose a large computational cost but produce diminishing returns with respect to 
the objective function. After some testing, we determined 
that a value of $\beta=0.9$ is appropriate for the Poisson error metric,
enabling more frequent outer updates for the overall GGN matrix.
We note that the choice of $\beta$ only 
affects the computational cost of the LM algorithm; the algorithm remains convergent as long 
as $\eta\in(0,1)$.

Finally, we note that there is no obvious indication in \Cref{alg:lm_inexact} that it 
is an \textit{AD-based matrix-free} implementation. Instead, this is implied in 
\Cref{ligne12} where the gradient and any matrix-vector products involving the Jacobian in
\Cref{eq:lma_magnitude} are computed ``on the fly'' with AD.

\subsection{An additional ``plug-in'' to deal with convex constraints}

In both of the canonical cases here, the SPR or the BPR problem (\Cref{sec:problems}), 
\Cref{alg:lm_inexact} is common to our LM approach. However,  an additional consideration is 
needed in the specific case of the BPR (\Cref{eq:bp}) problem because the sample $\OO$ and the 
probe $\PP$ updates need to be constrained to their respective convex sets $\setO$ 
and $\setP$. To achieve that aim, let us define the projection of the current 
estimate on the convex set of constraints $\mathcal{Z} :=\setO \cap \setP$ by
\begin{align}
	\Pi_{\mathcal{Z}}(\zzt) = \argmin_{x\in\mathcal{Z}} \norm{\zzt - x}^2.
	\label{eq:projection}
\end{align}
Following \cite{kanzow_jcam_2004} and \cite{fan_jimo_2013}, 
we can obtain a constrained and convergent LM update under 
these conditions  with the following strategy: (i) the unconstrained LM update from 
\Cref{alg:lm_inexact} is projected \via{} \Cref{eq:projection}, (ii) if this projected 
LM step does not decrease the error metric, and  if the search direction (at 
optimization iteration $t$) 
$\vec{s}_t = \Pi_\mathcal{Z}(\zzt_t + \Delta \zzt_t) - \zzt_t$ 
is a decrease direction for $f$ (Step 5 in \Cref{alg:projected_LM}), we perform a line search 
along $\vec{s}_t$ 
\cite[Section 4]{kanzow_jcam_2004}, 
(iii) if $\vec{s}_t$ is not a decrease direction, then we perform a standard projected 
gradient step. 
The ``line search'' step, with the search direction $\Delta \zzt_{ls,\,t}$, refers to 
a backtracking linesearch that calculates the step size $\alpha_t$ so that $\alpha_t$ 
satisfies the Armjio criterion:
\begin{align}
	\label{eq:armijo_criterion}
	f(\Pi_\mathcal{Z}(\zzt_t + \alpha_t \Delta \zzt_{ls,\,t})) \leq f(\zzt_t) - \sigma 
	\alpha_t \norm{\Delta \zzt_{ls,\,t}}^2
\end{align}
where $0 < \sigma \ll 1$. The ``projected gradient'' step refers to just the line search with 
the search direction  $\Delta \zzt_{ls,\,t} = 
-\grad f(\zzt)$.
 This strategy is detailed in \Cref{alg:projected_LM}. 

This simple projection strategy retains the pivotal assets of the LM approach: (i) that the 
iteration in \Cref{alg:projected_LM} is globally convergent, and (ii) that the algorithm can 
still preserve a superlinear convergence speed as long as the solution is not at the boundary of the constraint set 
\cite{kanzow_jcam_2004,fan_jimo_2013}.
\begin{algorithm}[H]
	\begin{algorithmic}[1]
		\Require $0 < \gamma \ll 1$, $0 < \tau_s \ll 1$ and $p_s > 1$.
		\Require Current guess $\zzt_t$, gradient $\grad{f(\zzt_t)}$.
		\Require Update direction $\Delta \zzt_t$ calculated using the \textit{unconstrained} 
		LM algorithm (\Cref{alg:lm_inexact}, Step 20).
		\If{$f(\Pi_\mathcal{Z}(\zzt_t + \Delta \zzt_t)) < \gamma f(\zzt_t)$}
		\State $\zzt_{t+1} = \zzt_t + \Delta \zzt_t$.
		\Else
		\State Set search direction $\vec{s}_t = \Pi_\mathcal{Z}(\zzt_t + \Delta \zzt_t) - 
		\zzt_t$.
		\If{$\grad f(\zzt_t)^T \vdot  \vec{s}_t \leq -\tau_s \norm{\vec{s}_t}^{p_s}$
		\State Calculate the step size $\alpha_t$ that satisfies the Armijo criterion in 
		\Cref{eq:armijo_criterion} for the search direction $\Delta \zzt_{ls,\,t} =\vec{s}_t$.
		\State Set $\zz_{t+1} = \zzt_t + \alpha_t \vec{s}_t$.
		\Else
		\State Calculate the step size $\alpha_t$ that satisfies the Armijo criterion in 
		\Cref{eq:armijo_criterion} for the search direction $\Delta \zzt_{ls,\,t} = -\grad 
		f(\zzt)$.
		\EndIf
	}
		\State Set $\zz_{t+1} = \zzt_t + \alpha_t \Delta \zzt_{ls,\,t}$.
		\EndIf
	\end{algorithmic}
	\caption{Additional projection step for the LM update.}
	\label{alg:projected_LM}
\end{algorithm}
In practice, since the probe and object constraints sets ($\setP$ and $\setO$ respectively) 
are separable, we can apply the projector in \Cref{eq:projection} just by performing the 
independent projections for $\PP$ and $\OO$.  

\subsection{Update schemes for the BPR problem}
\label{subsec:bpr_schemes}

For the BPR problem, the scheme outlined in \Cref{alg:lm_inexact,alg:projected_LM} details
a \textit{joint} optimization scheme which simultaneously optimizes both the sample $\OO$
and the probe $\PP$. In contrast, historical \cite{maiden_ultramic_2009,thibault_ultramic_2009} and also some popular modern \cite{fannjiang_arxiv_2020} BPR approaches update 
$\OO$ and $\PP$ in an alternating fashion. In \Cref{alg:alt_LM_ptycho} we also provide 
an alternate update LM scheme for the BPR problem.
\begin{algorithm}[H]
	\begin{algorithmic}[1]
		\Require Current guess $\zzt_t := (\tilde{\OO}_t$, $\tilde{\PP}_t)$. 
		\State Calculate the update $\Delta \tilde{\OO}_t$ using \Cref{alg:lm_inexact} (Step 
		4) with $\tilde{\PP}=\tilde{\PP}_t$ fixed.
		\State Set $\tilde{\OO}_{t+1} = \tilde{\OO}_t + \Delta \tilde{\OO}_t$.
		\State Calculate the update $\Delta \PP_t$ using \Cref{alg:lm_inexact} (Step 4) with 
		$\tilde{\OO}=\tilde{\OO}_{t+1}$ fixed.
		\State  Set $\zzt_{t+1} = (\tilde{\OO}_{t+1}, \tilde{\PP}_t + \Delta \tilde{\PP}_t)$.
	\end{algorithmic}
	\caption{Alternating LM update for the BPR problem}
	\label{alg:alt_LM_ptycho}
\end{algorithm}
Even though such an alternating scheme is often effective for first-order 
updates, it results in a loss of the sample-probe coupling information from the second-order 
curvature matrix; this could be detrimental to the speed of second-order minimization
algorithms. However, the sample and probe variables can have totally different 
scaling, and therefore the joint curvature matrix can be badly conditioned, while the separate 
sample-probe curvature matrices could still be individually well-conditioned. Consequently 
the PCG solution for the LM sub-problem of \Cref{eq:lma_magnitude} for the joint optimization 
scheme can have very slow convergence depending on the matrix conditioning. We can address 
this with appropriate preconditioning, as discussed below.

\subsection{Scaling and preconditioning with AD}
\label{subsec:lm_scaling_precond}

Practically speaking, AD frameworks dramatically simplify the implementation of any 
gradient-based iterative solver \cite{kandel_oe_2019}. For instance,  if we do not use either 
the scaling [\ie{} $\bm{D}=\bm{\mathcal{I}}$ in \Cref{eq:lma_magnitude}] or the CG 
preconditioning within the LM algorithm, then the AD-based implementation of 
\Cref{alg:lm_inexact} is totally agnostic to any analytical calculation whatsoever. 
Unfortunately, this is no longer the case if we use the scaling $\bm{D}=\Dg{\GG(\zzt)}$
and the CG preconditioner $\bm{P}=\Dg{\GG(\zzt) + \lambda \bm{D}}$. 
Actually, the required quantity $\Dg{\GG(\zzt)}$ is not a ``natural output'' of the AD 
framework so we derive it analytically (see \Cref{appendix:jacobi_lm}).

\section{Numerical Experiments}
\label{sec:experiments}

\begin{figure}[th]
	\includegraphics[width=\linewidth]{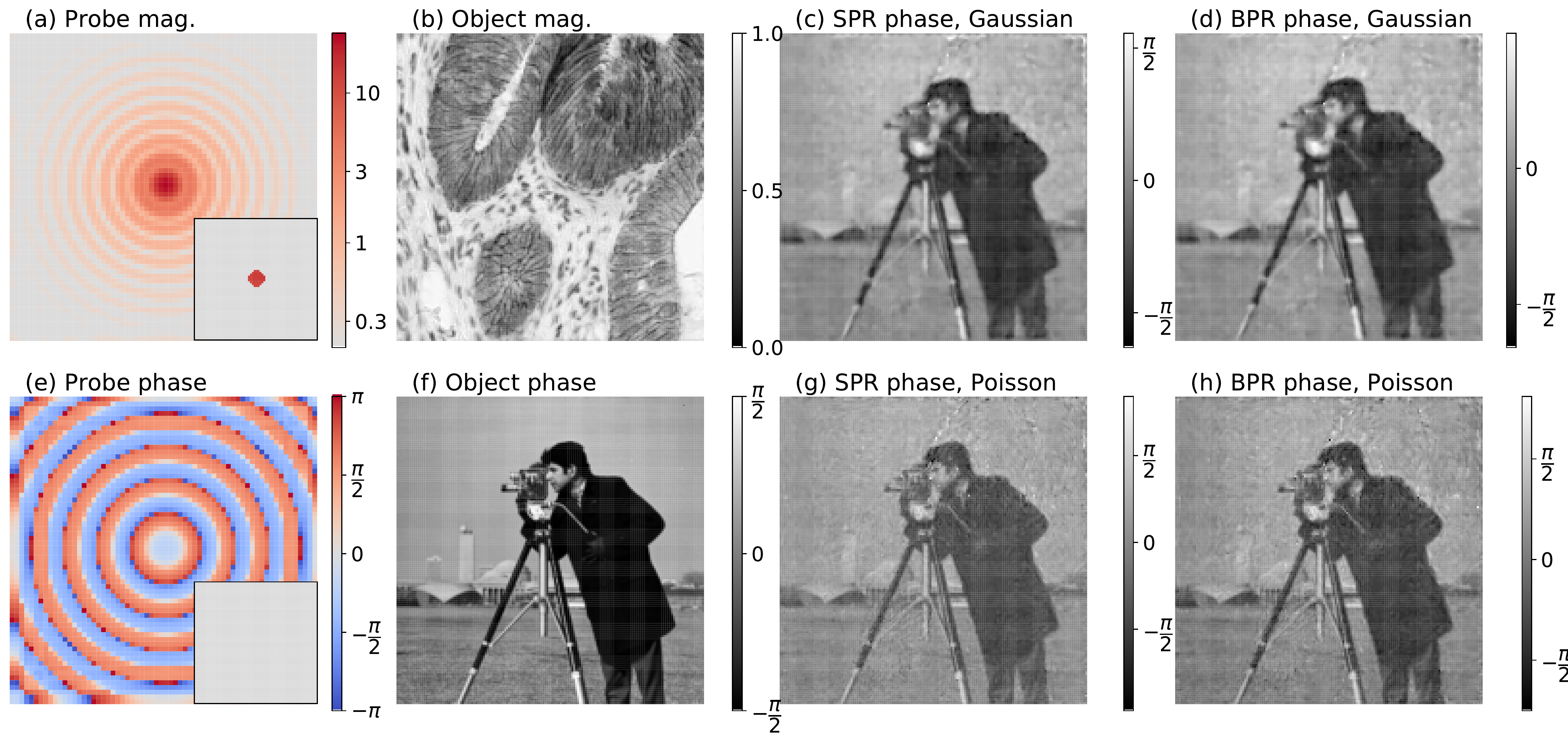}
	\caption{Simulated (true) probe (a) magnitude and (e) phase. The insets in (a) and (e) 
	show the initialization for the probe magnitude and phase respectively. Simulated (true) 
	object (b) magnitude and (f) phase. Reconstructed phase for the \flow{} SPR 
	setting with the (c) Gaussian and (g) Poisson error metric, using the PLM and PLM-S 
	algorithms respectively. Reconstructed phase for the \flow{} BPR case with the (d) 
	Gaussian and (h) Poisson error metric, using the PLM-J and PLM-J-S algorithms 
	respectively.}
	\label{fig:obj_probe}
\end{figure}

To test our proposed algorithms, we simulate a far-field transmission ptychography experiment 
and perform SPR and BPR with a variety of state-of-the-art reconstruction algorithms.
Our simulations use a $160\times 160$ pixels test object shown in \Cref{fig:obj_probe}(a,d) 
placed at the center of a $224\times 224$ pixel bounding box with the 
``bright-field'' boundary condition  applied \cite{fannjiang_ip_2019} (\ie{} with 
phase-less support set to $1.0$). We scan the object using a $64\times 64$ pixels probe 
generated by defocusing an Airy wavefront; the probe is shown in  \Cref{fig:obj_probe}(b,e). A ptychographic 
raster grid is then obtained by translating the probe latitudinally and longitudinally in 
steps of 5 pixels each, thus obtaining a dataset with a total of $1024$ 
noise-free intensity (diffraction) patterns. 

We also consider three different levels of Poisson counting noise depending on the integrated 
intensity of the probe: a ``low'' signal-to-noise ratio (SNR) setting with $10^3$ probe photons (with a fluence of $\flow=37.9$ photons per object pixel), a ``moderate'' SNR case with $10^4$ probe photons ($\fmod = 379$ photons/pixel), and a ``high'' SNR case with $10^6$ probe photons ($\fhigh=3.79\times 10^4$ photons/pixel). 
For all these simulation settings, we set a constant background level 
of $\bb_k=10^{-8}$ photons per object pixel.

With these datasets in hand, we attempt to solve the canonical problems 
defined in Sec.~\ref{sec:problems}. For each problem instance and SNR 
scenario, we generate five different uniformly random complex arrays 
with magnitude $\leq 1$ as the object initial-guess.
With each such object initialization, we run 1000 iterations of reconstruction.
For the (unconstrained) SPR problem, we use the true probe, and the algorithm only updates
the object guess. For the
(constrained) BPR problem which retrieves the probe as well, we use 
the constraint sets
\begin{align}
	\setO = \{\OO\in\Complexes^N : \abs{\OO}\leq 1\};\quad 
	\setP = \{\PP\in\Complexes^M : \abs{\PP}\leq 10^8\}.
	\label{eq:constraints}
\end{align}
The initial guess for the probe wavefront is a single-mode phaseless 
circular aperture with its width equal to the central lobe of the Airy
beam (before defocus), and with the integrated intensity set to that
of true probe wavefront, as shown in \Cref{fig:obj_probe}(c,f). 

At the end of every iteration, we assess the progress made by the optimization
process: we first use the subpixel registration algorithm \cite{guizar_josaa_2004} to 
calculate the normalized object reconstruction error from the ground truth, and then estimate the 
mean object reconstruction error $\oerr_t$ (where $t$ indexes the iteration number) averaging 
the error from the five independant object initializations.  For the BPR problem, we also 
record the mean probe reconstruction error $\perr_t$. 

We estimate the evolution of the computational cost of the optimization process
by tracking the number of floating point operations (flops) required per iteration of the
algorithm. For example, for the LM algorithm, we estimate the computational cost for a given 
iteration by tracking: a) the number of $\lambda$ updates required in this iteration, b) the 
number of CG iterations required for each such $\lambda$ update, and c) the number
of projected gradient line search iterations required in this iteration. We separately 
estimate the number of flops required for each CG iteration, for any extra computation 
required (outside of the inner CG loop) at each $\lambda$ update, and for each iteration of the
projected gradient linesearch. By combining these appropriately, we get the overall 
computational cost for the iteration.

Finally, to define a convergence indicator for the optimization procedure, we use 
a sliding window over the time series for $\oerr_t$ to calculate the root mean square value
\begin{align}
	\mbox{WindowRMSD}_j = \sqrt{\frac{\sum_{w=0}^{W} (\oerr_w - \bar{\oerr}_j)^2}{W - 1}}
	\label{eq:window_rmsd}
\end{align}
where $W=100$ is the window size, $j$ indexes the current window, and $\bar{\oerr}_j$ is the 
mean of the values in the current window. The optimization procedure can then be said to have 
``converged'' at iteration $j$ if 
\begin{align}
	\mbox{WindowRMSD}_j \le \epsilon_c 
	\quad \text{and}\quad 
	\bar{\oerr}_j \leq  \bar{\oerr}_{j'} \quad 
	\text{for all } j'\neq j,
	\label{eq:window_rmsd_conv}
\end{align}
where $0 < \epsilon_c \ll1$ is a constant. For the subsequent analyses, we define the point 
of convergence using with $\epsilon_c=10^{-3}$, $\epsilon_c=2\times 10^{-3}$, 
and $\epsilon_c=3\times 10^{-3}$ for the \fhigh{}, \fmod{} and \flow{} cases respectively. 

In the following numerical experiments, all our reconstruction algorithms use 
single precision floating points for the numerical calculations. 
For the LM calculations, we use the recommendations in 
\cite{kanzow_jcam_2004,fan_jimo_2013,huang_jocm_2017} and set 
$\mu_0=10^{-5}$, $\mu_{min}=10^{-8}$, $\nu=1$, $\rho_{min}=10^{-4}$, 
and $\kappa=4$ in \Cref{alg:lm_inexact}, $\sigma=10^{-4}$ in \Cref{eq:armijo_criterion}, 
and $\gamma=10^{-6}$, $\tau_s=10^{-8}$, and $p_s=2.1$ in \Cref{alg:projected_LM}.

\subsection{Performance of the matrix-free LM solver for the SPR problem}
In this subsection, we present the performance analyses of the LM solvers applied to the SPR 
problem with the Gaussian and Poisson error metrics. Here, we compare the basic truncated LM 
algorithm (denoted as \textbf{LM}) and the LM algorithm implemented with the diagonal 
preconditioning and scaling (denoted as \textbf{PLM}), with the following state-of-the-art 
first-order algorithms (described in \Cref{appendix:first_order_algs}): 
\begin{itemize}
	\item \textbf{NCG/PNCG}: the nonlinear conjugate gradient (NCG) algorithm and the 
	preconditioned NCG algorithm for both the Gaussian and Poisson error metrics.
	\item \textbf{NAG}: Nesterov's accelerated gradient method for the Gaussian error 
	metric. 
\end{itemize}
\subsubsection{The Gaussian error metric}
\begin{figure}[th]
\includegraphics[width=\linewidth]{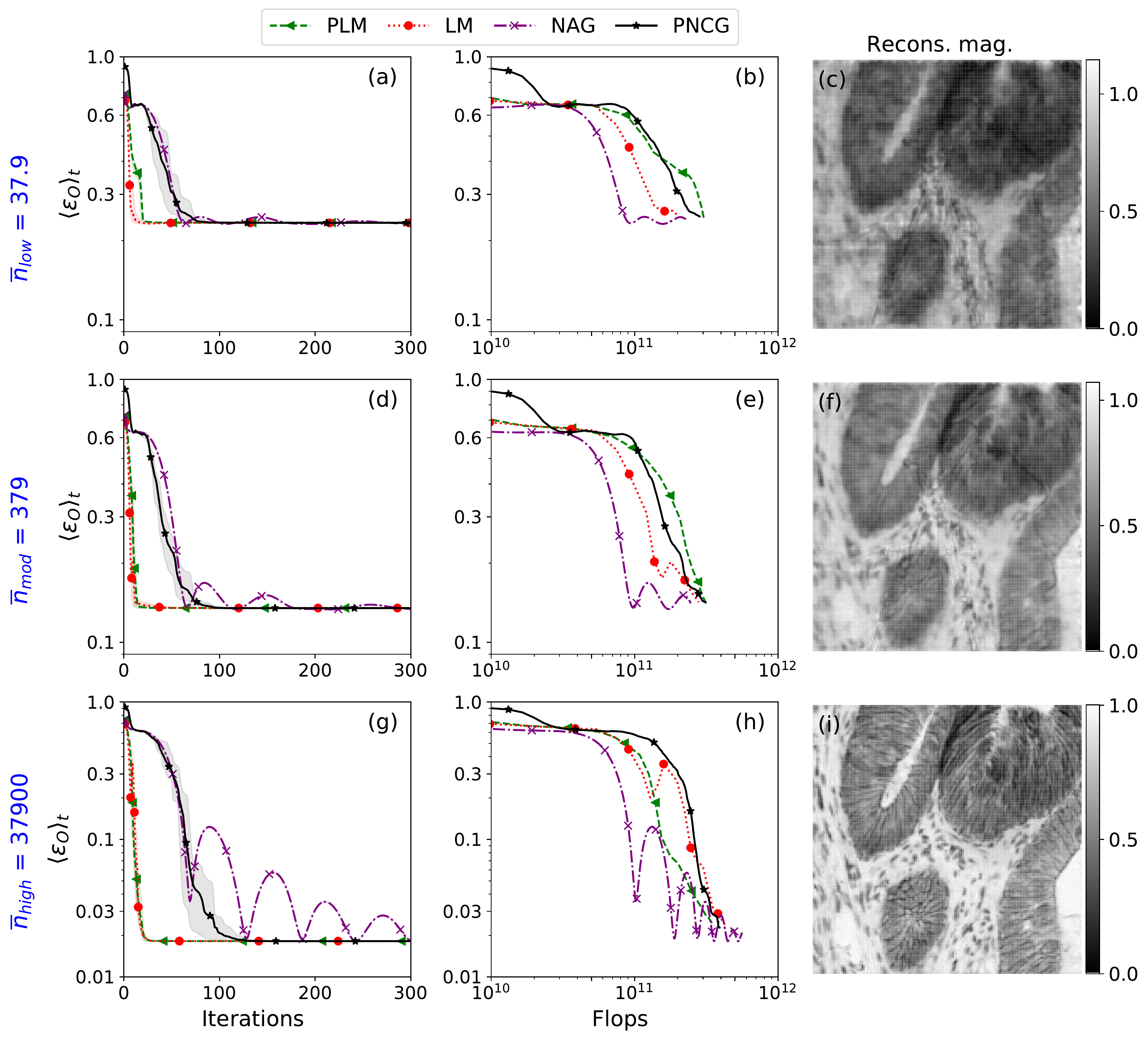}
\caption{Standard ptychographic reconstructions (SPR) of the object with a known probe using the Gaussian error metric. The top, middle, and bottom rows 
display the results for the \flow{}, \fmod{}, and \fhigh{} cases respectively. Subplots (a,d,g) display the	history 
of the normalized object reconstruction error \oerr{} over the first 300 optimization iterations,
subplots (b,e,h) display the history of \oerr{}
\textit{up to convergence} as a function of the computational cost in flops, and 
subplots (c,f,i) display the reconstructed object magnitude ($\abs{\langle \OO\rangle}$) 
obtained from the PLM algorithm.}
\label{fig:spr_gaussian}
\end{figure}
In \Cref{fig:spr_gaussian}, we demonstrate the performance of the LM method for the SPR 
problem with the Gaussian error metric. Here, subplots (a, d, g) clearly show that, for all 
the noise levels, 
the LM algorithms converge to the solution within a few iterations,
 thus displaying a convergence rate significantly faster than any of the first-order 
 algorithms tested.In fact, these subplots indicate that the LM iterates
have the expected superlinear rate of convergence (\ie{} the error decreases at a 
faster-than-linear rate in the semilog plot), but this is difficult to actually verify 
numerically due to the limited dynamic range of the reconstruction problems.

Subplots (b, e, h) show that, in terms of the real computational cost (in flops), for the 
\flow{} setting, the NAG algorithm reaches the solution much faster than the other 
methods. However, as we increase the SNR, this advantage decreases, such that at the \fhigh{} 
setting, the computational cost of the LM and PLM methods are comparable to that of the NAG 
and PNCG methods.

Subplots (c, f, i) present the recovered object magnitudes for the PLM algorithm at the 
various noise levels;the figures demonstrate that the object recovered
in the the high SNR setting (where $\oerr=0.025$) is difficult to distinguish from the true 
object, but the moderate and low SNR settings (where $\oerr=0.15$ and $\oerr=0.25$ 
respectively) show a clear deterioration in the object recovered.

Another point of interest is that the PLM algorithm 
 does not necessarily provide an improvement in the computational cost over the basic LM method.
 This result is in keeping with the observation \cite{qian_ipa_2014} 
 that the SPR problem is generally well-conditioned as long as there is a sufficient overlap between  adjacent probe 
 positions. In this case, the analytical calculation for 
 the preconditioner is not strictly necessary, 
and the basic LM algorithm typically suffices. 
We present the full quantitative results in \Cref{tab:gaussian_spr}.

\subsubsection{The Poisson error metric}
\label{subsec:spr_poisson}
\begin{figure}[th]
	\includegraphics[width=\linewidth]{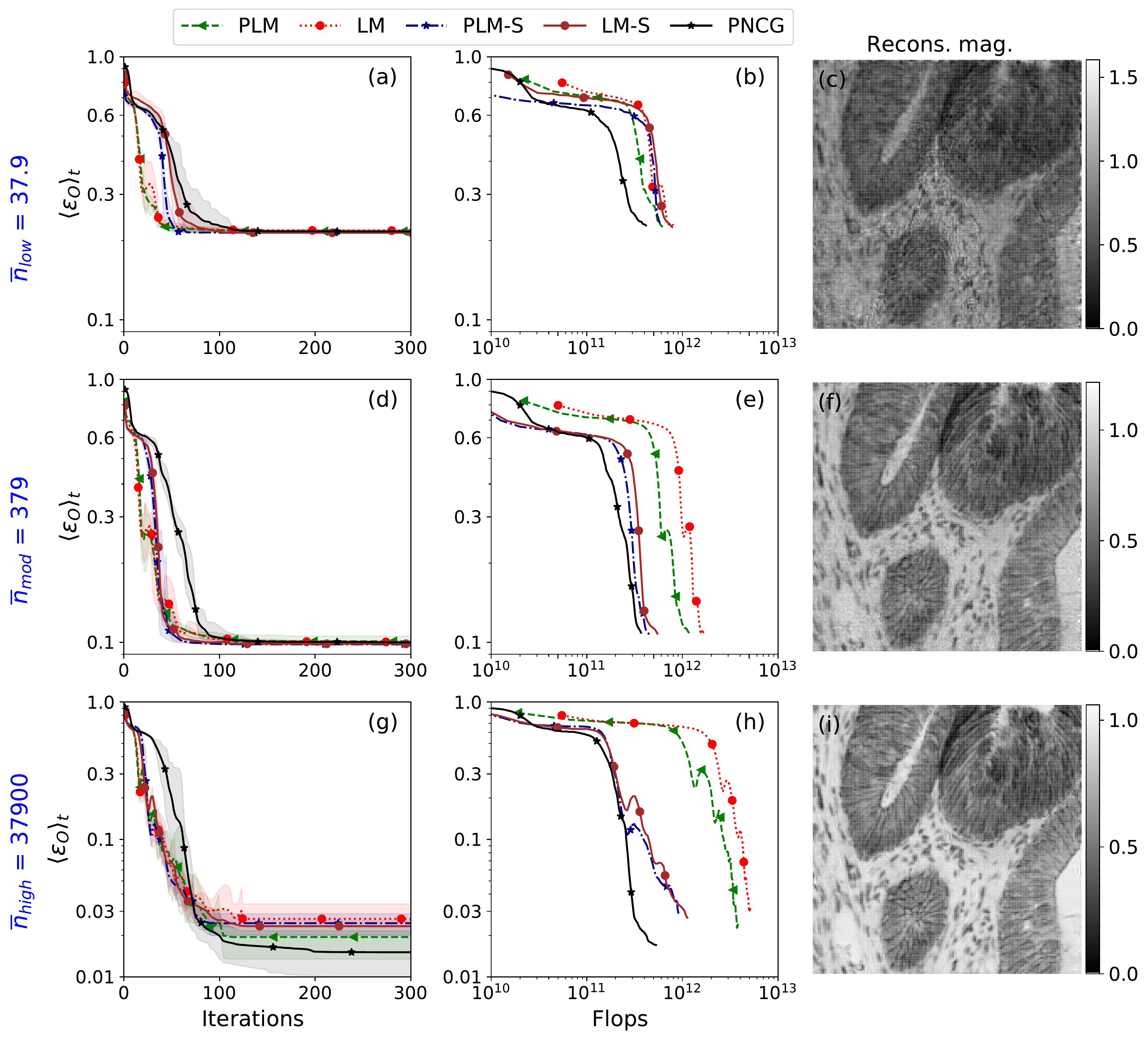}
	\caption{Standard ptychographic reconstructions (SPR) of the object with a known probe using the Poisson error metric.  The notations LM-S and PLM-S denote the respective LM method optimizing the 
	surrogate formulation of the Poisson error metric. 
	As with \Cref{fig:spr_gaussian}, subplots (a, d, g) and (b, e, h) show the history of \oerr{} 
	with the iterations and flops (to convergence) respectively.
	Subplots (c, f, i) display the reconstructed object magnitude obtained from the PLM-S 
	algorithm. }
	\label{fig:spr_poisson}
\end{figure}
In \Cref{fig:spr_poisson} we present the reconstruction results for the SPR problem with the 
Poisson error metric.  In  subplots (b, e, h), the lines for the LM (red dots) and PLM (green 
dashes) algorithms clearly show that the individual LM updates incur a very large 
computational cost, particularly at the initial stages of the optimization procedure, and that this effect becomes more significant as we increase the dynamic range of the problem. This 
result stands in stark contrast to the results for the Gaussian error metric 
(\Cref{fig:spr_gaussian}). We examine this discrepancy in \Cref{appendix:ggn_efficacy}, where 
we show that, at points far away from the minimum, the GGN approximation does not effectively 
capture the curvature of the Poisson error metric. This result suggests that the PNCG 
algorithm could be a much more effective choice if we want to minimize the Poisson model for 
the unconstrained SPR problem.

For practical use cases (even for the SPR case) we often have experimental 
constraints (such as the object constraint in \Cref{eq:constraints}) that we want to apply during 
the optimization procedure, and the basic PNCG algorithm is no longer a robust option. In such 
cases, we can perform fast LM optimization by using a surrogate formulation of the Poisson error 
metric. We can obtain this surrogate formulation by redefining the expected magnitude (from 
\Cref{eq:expected_count,eq:magnitude}) as:
\begin{align}
	\bzeta^2 \,=\, \abs{\hbpsi_k}^2 + \bb_k + \varsigma_t
\end{align}
where we monotonically decrease $\varsigma_t$ from a large stabilizing value to $0$ as we 
proceed with the optimization iteration $t$, so that we revert to the true Poisson metric after
a predefined number of optimization iterations (see details in 
\Cref{appendix:poisson_surrogate}). 
We introduce the notation \textbf{LM-S} and \textbf{PLM-S} 
to denote the LM algorithms optimizing this surrogate model\footnote{In our numerical tests (not 
shown), we find that the first-order methods tested do not show any appreciable acceleration 
when we use the surrogate formulation of the Poisson error metric.}. 
The results in \Cref{fig:spr_poisson} (and \Cref{tab:poisson_spr}) show that this 
simple reformulation significantly reduces the computational cost 
required per LM iteration at the initial states of the
optimization (when we are far away from the minimum) for both the moderate SNR and high SNR settings; the LM optimization cost is now only slightly higher than that for the PNCG method. Conversely, our use of the surrogate Poisson formulation and a loose acceptance criterion for the CG substep (with $\beta=0.9$ in \Cref{eq:lma_termination_eta}) has the consequence that the LM iterates are no longer superlinearly convergent; they nonetheless converge faster than the PNCG algorithm for the \flow{} and \fmod{} cases.

We observe in subplots (c,f) that the objects recovered  for the \flow{} and \fmod{} cases with the PLM-S method (with $\oerr=0.24$ and $\oerr=0.11$ respectively) look sharper  than that obtained with the Gaussian error 
metric, but the range of magnitudes obtained is much larger for the Poisson error metric. For the \fhigh{} case, optimizing either of the error metrics results in similarly accurate reconstructions (see \Cref{tab:gaussian_spr,tab:poisson_spr}). We note that the Gaussian metric can be interpreted as a consistent approximation of the Poisson metric for the high SNR regime.
In this regime, the Gaussian error metric can be optimized with a lower computational effort, and it is not clear if we actually want to optimize the Poisson metric at all.

\subsection{Performance of the matrix-free LM solver for the BPR problem}
In this subsection, we present the performance analyses of the LM solvers applied to the BPR 
problem with the Gaussian and Poisson error metrics. In the numerical comparisons, we test the 
following LM variations:
\begin{itemize}
	\item \textbf{LM-A}: the basic alternating LM algorithm, as described in 
	\Cref{alg:alt_LM_ptycho}.
	\item \textbf{PLM-A}: the LM-A algorithms modified to use diagonal scaling and 
	preconditioning.
	\item \textbf{PLM-J}: the joint optimization scheme that simultaneously updates the
	$\OO$ and $\PP$ for the BPR problem.
\end{itemize}
Here, we do not include the results for the basic joint optimization scheme (without 
preconditioning and scaling) since these reconstructions \textit{do not} converge within 1000 
iterations. We compare these to the following state-of-the-art first order BPR methods 
(described in \Cref{appendix:first_order_algs}):
\begin{itemize}
	\item \textbf{PHeBIE}: the proximal heterogeneous block implicit-explicit (PHeBIE) method 
	\cite{hesse_siamjis_2015}.
	\item \textbf{ADMM}: the alternating directions minimization (ADMM) method
	\cite{chang_sjis_2019}. 
	\item \textbf{ePIE}: the extended Ptychographic Engine (ePIE) method 
	\cite{maiden_ultramic_2009}.
\end{itemize}
Unlike the PHeBIE and ADMM methods, which are provably convergent for the BPR problem, the ePIE method does not provide convergent updates \cite{godard_oe_2012,hesse_siamjis_2015}. Nevertheless, the ePIE method is still widely used in the community and we therefore include the results here.

\subsubsection{The Gaussian error metric}
\label{subsubsec:gaussian_bpr}
\begin{figure}[th]
	\includegraphics[width=\linewidth]{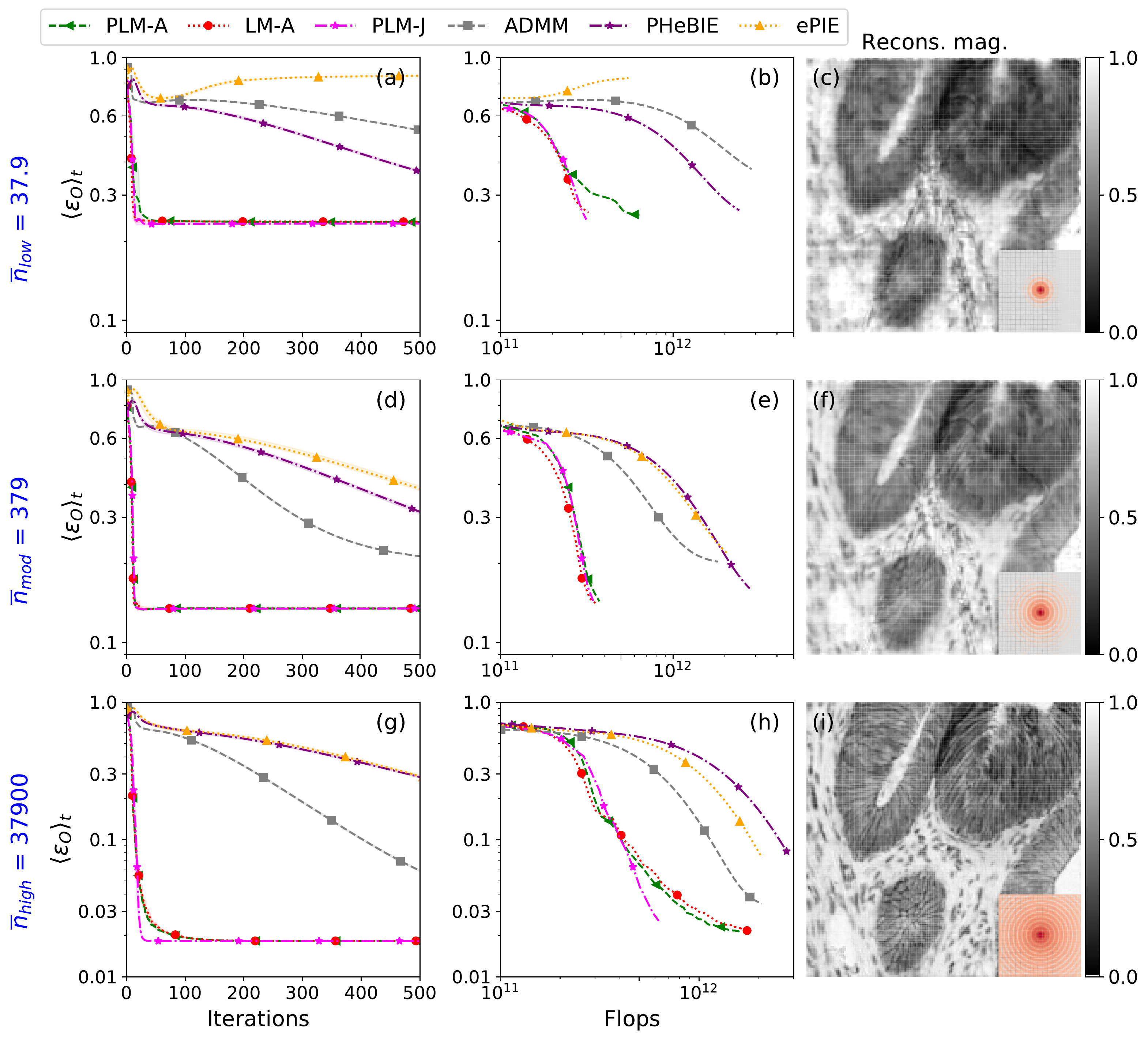}
	\caption{Blind ptychographic reconstructions (BPR) of the object and the probe with the Gaussian error metric. 
	As with \Cref{fig:spr_gaussian}, subplots (a, d, g) and (b, e, h) show the history of \oerr{} 
	with the iterations and flops (to convergence) respectively. Subplots (c, f, i) display the reconstructed 
		object magnitude and probe magnitude (in the inset) obtained from the PLM-J algorithm.}
	\label{fig:bpr_gaussian}
\end{figure}
In \Cref{fig:bpr_gaussian}, we present the performance results for the BPR problem with the Gaussian error metric. 
The joint (PLM-J) optimization procedure
converges to the minimum in significantly fewer iterations, and with lower computational effort, than any of the first-order 
algorithms. Again, \Cref{fig:bpr_gaussian}(g) seems to indicate that the PLM-J algorithm is superlinearly convergent. Meanwhile, the convergence rate of the alternating (LM-A and PLM-A) optimization 
procedures decreases as we increase the dynamic range of the problem, which shows that the 
sample-probe coupling information in the second-order curvature matrix becomes more
important for problems with higher SNR. Another point of interest is that the LM-A 
algorithm does not benefit appreciably from the application of preconditioning,
which indicates that the object and probe sub-problems are individually well-conditioned;
however, the joint optimization approach is not achievable without the application of a 
 preconditioner. 
 
Among the first-order methods tested, the ePIE algorithm, which does not use 
any constraints, does not improve the object reconstruction at all in the \flow{} case. More 
generally, the ePIE and the PHeBIE algorithms show slow convergence in all the settings 
tested. 
Meanwhile, the ADMM convergence depends greatly on the choice of the the penalty parameter 
$\varrho$ (see \Cref{appendix:admm}). In our experience, tuning $\varrho$ is a computationally 
expensive procedure, which becomes progressively more difficult as we lower the SNR. This 
effect is evident in subplots (b, e, h): the ADMM reconstructions are relatively (compared to 
the LM methods) more expensive and of lower quality as we move from bottom to top (see also 
\Cref{tab:gaussian_bpr}). It is likely that we can further tune the value of $\varrho$ to 
improve the convergence rate and solution quality, but this would require even more 
computational effort and is beyond the scope of this paper.

We present the full results for the converged algorithms in \Cref{tab:gaussian_bpr}.

\subsubsection{The Poisson error metric}
\begin{figure}[th]
\includegraphics[width=\linewidth]{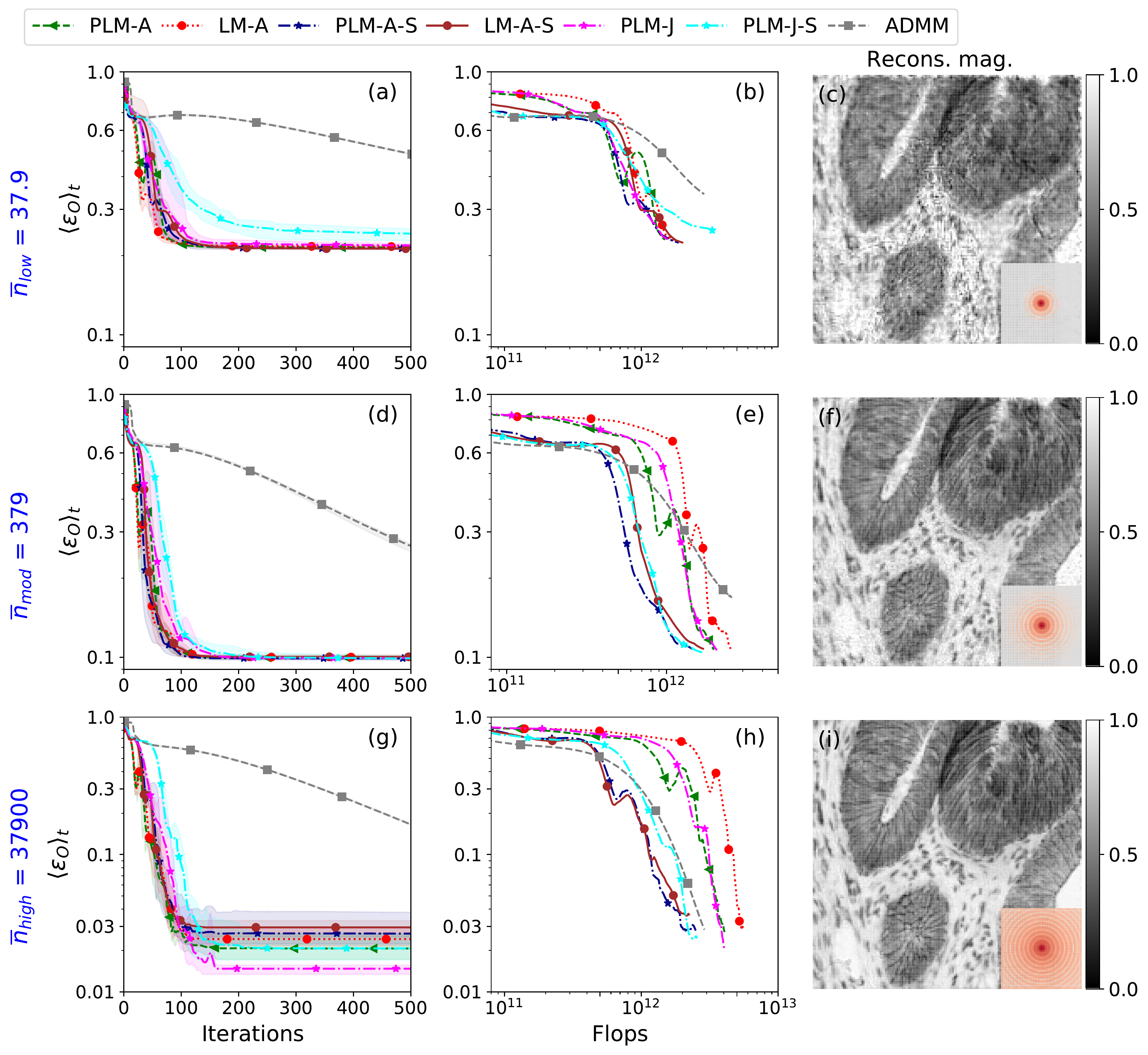}
	\caption{Blind ptychographic reconstructions (BPR) of the object and probe with the Poisson error metrics,
		with the notations LM-A-S, PLM-A-S, and PLM-J-S denoting the respective
		 LM methods for the surrogate formulation of the  Poisson metric.
		 As with \Cref{fig:spr_gaussian}, subplots (a, d, g) and (b, e, h) show the history of \oerr{} 
		 with the iterations and flops (to convergence) respectively.
		 Subplots (c, f, g) display the reconstructed object magnitude and probe magnitude (inset) obtained from the PLM-J-S algorithm.
	}
	\label{fig:bpr_poisson}
\end{figure}
Figure \ref{fig:bpr_poisson} shows the reconstructions for the for the various SNR cases with 
the Poisson error metric. These results again show that the surrogate 
formulation of the Poisson error metric reduces the computational cost required for the 
optimization procedure for the \fmod{} and \fhigh{} settings. These results differ from that 
in \Cref{subsubsec:gaussian_bpr} in two ways: i) the LM  iterates (even with the surrogate 
formulation) have a slower rate of convergence, and ii) optimization with the PLM-J algorithm 
requires a computational effort similar to that needed for the PLM-A algorithm. We can 
attribute both of these observations to the fact that we use a loose acceptance criterion for 
the CG substep when optimizinig the Poisson error metric. However, it is possible that the 
PLM-J algorithm would provide an improvement over the PLM-A algorithm for a BPR problem with a 
much larger dynamic range.  Finally, the ADMM method again displays the behavior described in 
\Cref{subsubsec:gaussian_bpr}.

Similar to the SPR case, the PLM-J-S reconstruction for the \flow{} and \fmod{} case (with $\oerr=0.25$ and $\oerr=0.11$ respectively) is sharper than the reconstruction for the Gaussian error metric.
For the \fhigh{} setting, the caveat described in the SPR case (\Cref{subsec:spr_poisson}) again 
applies (see \Cref{tab:poisson_bpr}).

\section{Discussion}
\label{sec:discussion}

The comparisons of the various approaches discussed above were all drawn from a relatively simple simulated case. As we summarize below, this provided us the opportunity to delve deeply into the regimes of relative success and failure of a wide range of phase retrieval approaches, including the LM-based implementations that we developed. It is also important to note that when imagining use of LM algorithms for ptychography experiments as well as broader general use in phase retrieval, certain additional aspects should be considered as well, which we touch on below. 

Through our experiments, we find that the LM algorithms shine in applications with the 
Gaussian error metric, both for the SPR and BPR problems. For the SPR problem, at all the 
noise levels examined, the preconditioned LM algorithm is found to converge superlinearly, in significantly fewer iterations than 
any of the first-order algorithms. For the BPR problem, the joint optimization (PLM-J) method 
shows similarly excellent performance. Remarkably, in all these cases, the LM algorithms have 
a true computational cost that is comparable to, or even lower than, that of the 
best-performing first-order algorithm with the
  additional advantage of minimal hyperparameter tuning.

The Poisson error metric is more difficult to optimize (in a computationally efficient manner) with the LM algorithm, and requires the use of a surrogate, stabilized, formulation of the Poisson error metric as well as a loose acceptance criterion for the CG substep. These modifications reduce the convergence rate so that it is no longer superlinear, but increase the computational efficiency of the LM method. For the unconstrained SPR problem, this modified LM algorithm converges at a rate  similar to the first-order PNCG method, but has a slightly higher computational cost. For applications with constraints, however, the (modified) LM method remains the most performant choice. Notably, for high SNR experiments, the Gaussian error metric
is generally expected to serve as a robust proxy for the Poisson error metric and produce 
similar reconstruction results.  Under these conditions, the LM methods perform as well as 
the PNCG method even for unconstrained problems.

By incorporating the projected gradient method, 
the LM algorithm provides convergence guarantees in both the unconstrained and constrained 
optimization settings with the benefit of minimal parameter tuning. Looking at comparable approaches, even though the PHeBIE and ADMM methods both guarantee convergence for 
BPR problems with constraints, and can be easily modified to also apply to constrained SPR problems, they can be difficult to accelerate. The PHeBIE algorithm is a steepest 
descent procedure that depends on the use of small, stable, step sizes for its convergence; 
accelerating this method requires a sophisticated domain decomposition approach \cite{hesse_siamjis_2015}. The ADMM algorithm, on the other hand, 
permits bigger step sizes \cite{chang_sjis_2019}, but only with a careful choice of the ADMM 
penalty parameter, the value of which often changes from problem to problem (see 
\Cref{fig:admm_varrho}). On the other hand, the projected LM procedure generally works with minimal modifications in all these different settings.

The fact that the second-order LM algorithm has a  computational cost \emph{comparable} to first-order methods is itself quite remarkable, and further improvements and broader applicability can easily be imagined. 
The primary reason for computational parity of these methods is that our LM implementation uses the reverse-mode AD framework to avoid the construction of the full GGN matrix (with $N^2$ elements); the only drawback of this approach is a higher memory cost, about $4\times$ that for a first-order method. Additionally,
our work here also showcases the GGN-based extension of the classical LM algorithm, which can potentially be applied beyond ptychography to 
problems with general convex error metrics. Among the  implications of this is that 
the LM procedure can be used for phase retrieval problems with other noise statistics 
(\eg{} a mixed Gaussian-Poisson noise setting \cite{jezierska_espc_2011}), other error metrics (\eg{} for robust 
regression models), or even for entirely different classes of optimization problems.

To determine the ease of generalization of the LM method we need an additional consideration: whether a preconditioning strategy is necessary for LM optimization. For the SPR problem, the LM procedure has similar performance both with and without any preconditioning as long as there is sufficient overlap between adjacent probe positions. Likewise, the alternating minimization strategy for the BPR problem is also not dependent on the use of a preconditioner. However, the BPR joint optimization strategy, which is faster than the alternating optimization strategy for higher SNR regimes, requires the use of a preconditioner. On the one hand, 
typical phase retrieval problems (in other experimental modalities) only attempt to retrieve the illuminated object, and are similar to the SPR problem in this regard. This suggests that for general \textit{well-conditioned} phase retrieval problems, a preconditioner is not strictly necessary. In this scenario, the LM algorithm is entirely AD-based, with no analytical calculations required, and can thus be used in a drop-in fashion to solve these problems. On the other hand, if our phase retrieval problem is \textit{ill-conditioned}, or if we desire a  BPR-like joint optimization strategy, then we have to rely on a preconditioner. In this case, one option is to use the diagonal elements of the GGN matrix ($\Dg{\GG(\zzt)}$) for the preconditioning; we provide the analytical expressions for these for the BPR problem and they are straightforward to modify for other phase retrieval applications\footnote{We can even use these expressions to develop preconditioned first-order methods, such as the PNCG method (which we use here) or ``scaled gradient'' optimization methods (the Accelerated Wirtinger Flow 
\cite{xu_arxiv_2018} uses such an approach for the Gaussian error metric).}. When this analytical calculation is not feasible, however, we can instead use stochastic matrix-free techniques to compute an unbiased estimate of $\Dg{\GG(\zzt)}$ in a computationally efficient manner through matrix-vector products alone  
\cite{bekas_anm_2007,chapelle_nips_2011,martens_phd_2016} (for
example, the recently published 
AdaHessian algorithm uses such a scheme for mini-batch second-order GGN optimization 
\cite{yao_arxiv_2020}). We expect that these modifications can allow for use of the LM method for general SPR-like and BPR-like phase retrieval problems.

\section*{Funding}

This work was supported by the U.S. Department of Energy, Office of Science, Basic Energy 
Sciences, Materials Science and Engineering Division.
CJ acknowledges partial support by the Advanced Photon Source, a U.S. Department of Energy 
(DOE) Office of Science User Facility operated for the DOE Office of Science by Argonne 
National Laboratory under Contract No. DE-AC02-06CH11357 and by the National Institutes of 
Health under R01 GM104530, and R01 MH115265.
YN was partially supported by U.S. DOE Contract No. DE-AC02-76SF00515.

The submitted manuscript has been created by UChicago Argonne, LLC, Operator of Argonne 
National Laboratory (“Argonne”). Argonne, a U.S. Department of Energy Office of Science 
laboratory, is operated under Contract No. DE-AC02-06CH11357. The U.S. Government retains for 
itself, and others acting on its behalf, a paid-up nonexclusive, irrevocable worldwide license 
in said article to reproduce, prepare derivative works, distribute copies to the public, and 
perform publicly and display publicly, by or on behalf of the Government. The Department of 
Energy will provide public access to these results of federally sponsored research in 
accordance with the DOE Public Access Plan. http://energy.gov/downloads/doe-public-access-plan

\clearpage

\appendix

\section{Using reverse-mode AD for matrix-free GGN-vector products}
\label{appendix:AD_mvp}

In this appendix, we describe the mechanism through which we use the reverse-mode AD framework
to calculate the linear system (in \Cref{eq:lma_magnitude}) that defines a single iteration of 
the LM algorithm. 
We can rewrite this linear system as
\begin{align}
	\GG(\zzt) \vdot  \Delta \zzt_\star + \lambda \II \vdot \Delta \zzt_\star  
	= -\JJ(\zzt)^T \vdot \grad_{\zeta}\ell(\bzeta).
	\label{eq:ggn_magnitude_appendix}
\end{align}
where $\JJ(\zzt)\equiv\JJ_\zeta(\zzt)$ and $\GG(\zzt)\equiv\GG^\ell(\zzt)$ 
are respectively the Jacobian and GGN matrices associated with the magnitude-based error  
metric $\ell$ defined in \Cref{eq:ell_magnitude}. A further rewriting of 
\Cref{eq:ggn_magnitude_appendix} leads to the equation
\begin{align}
	\overbrace{\JJ^T (\zzt) \vdot 
		\underbrace{( \grad^2_{\bzeta} \ell(\bzeta) \vdot 
			\overbrace{( \JJ(\zzt) \vdot \Delta \zzt_\star)}^{\jvp{}})}_{\hvp{}}}^{\jtvp{}}
	+ \lambda \II \vdot \Delta \zzt_\star  
	= -\overbrace{\JJ(\zzt)^T \vdot 
		\grad_{\zeta}\ell(\bzeta)}^{\jtvp{}}.
	\label{eq:ggn_magnitude_appendix_2}
\end{align}
Thus, to formulate a \textit{matrix-free} LM procedure, 
we need the ability to calculate the \textit{Jacobian-vector product} (\jvp{}), the 
\textit{Jacobian-transpose-vector product} 
(\jtvp{}), and the \textit{Hessian-vector product} (\hvp{}) in a matrix-free fashion.

To illustrate the mechanism for these matrix-vector product calculations, we
define a convenient abstraction $\Gamma(\zzt) = \bzeta$ and assume that 
$\Gamma$ can be expressed as a composition of $L$ transformations such that
\begin{align}
	\Gamma = \Gamma_L \circ \Gamma_{L-1} \circ\dots\circ \Gamma_1,
	\label{eq:gamma_composition}
\end{align}
where $\Gamma_1(\zzt)= \bzeta_1,\dots,\Gamma_{L-1}(\bzeta_{L-2}) = \bzeta_{L-1}$ are the 
intermediate 
outputs obtained when we calculate $\Gamma(\zzt) = \Gamma_L(\bzeta_{L-1}) = \bzeta$.
In practice, it is easy to associate these intermediate abstractions with the actual elemental
functions that we use to calculate $\bzeta$ 
(\Cref{eq:exit_wave,eq:expected_count,eq:magnitude}).
However, by using these abstractions we want to emphasize that the mechanism described in 
this section
can be used similarly in general multivariable optimization problems beyond 
just the ptychography applications.

First, we apply the chain rule of differentiation to the function composition in 
\Cref{eq:gamma_composition} and take the transpose to get the identity
\begin{align}
	\JJ_\Gamma(\zzt)^T &= \JJ_{\Gamma_1}(\zzt)^T \vdot
	\dots \vdot
	\JJ_{\Gamma_{L}}(\hh_{L-1})^T,
	\label{eq:multivar_chain_rule_transpose}
\end{align}
where $\JJ_\Gamma,\,\JJ_{\Gamma_L},\dots,\,\JJ_{\Gamma_1}$ are the Jacobian matrices for the 
functions $\Gamma,\,\Gamma_L,\,\dots,\,\Gamma_1$ and are defined as usual. We can then 
calculate the \jtvp{} of the Jacobian $\JJ_\Gamma$
and an arbitrary vector $\omega\in\Reals^m$ by simply following the sequence of 
matrix-vector products in the order indicated by the brackets in 
\begin{align}
	\JJ_{\Gamma}(\zzt)^T \vdot \bomega = \JJ_{\Gamma_1}(\zzt)^T \left[
	\vdot \left[
	\dots \vdot \left[
	\JJ_{\Gamma_{L}}(\hh_{L-1})^T \vdot \bomega \right]
	\dots \right]
	\right].
	\label{eq:arbitrary_jtvp_sequence}
\end{align}
We thus have a powerful mechanism for matrix-free calculations of \jtvp{}s
for function compositions of any length and complexity: 
if we have a method to calculate the \jtvp{} of an arbitrary 
$(\mbox{Jacobian},\, \mbox{vector})$ pair in a matrix-free fashion, 
we can apply this same method repeatedly to get the desired $\jtvp{}$ for any desired function 
composition. 
This mechanism forms the basis of the reverse-mode AD procedure and is referred to as the 
``adjoint'' model in automatic differentiation 
\cite{griewank_siam_2008}.

In contrast to the \jtvp{} operation, the \jvp{} operation is not a natural output of the 
reverse-mode AD
procedure, and instead requires a composition of two separate \jtvp{} operations 
\cite{townsend_blog_2017}. 
To illustrate this procedure, we first represent the \jtvp{} calculation from 
\Cref{eq:arbitrary_jtvp_sequence} as a function $\xi$ which has the action:
\begin{align}
	\xi(\zzt, \bomega) = \JJ_{\Gamma}(\zzt)^T \vdot \bomega.
	\label{eq:jvp_step_1}
\end{align}
For $\zzt=\zzt_0$ fixed, $\xi(\zzt_0, \bomega)=\xi(\omega)$ 
is just a linear transformation of $\bomega$. 
Hence, the Jacobian of $\xi$ with respect to $\bomega$ is exactly the matrix  
$\JJ_{\Gamma}(\zzt_0)^T$ that does not depend on the actual value of $\bomega$:
\begin{align}
	\JJ_\xi(\bomega) = \JJ_{\Gamma}(\zzt_0)^T.
	\label{eq:jvp_step_2}
\end{align}
Now, for an arbitrary $\vv\in \Reals^{2n}$,  we can calculate the \jtvp{}
\begin{align}
	\JJ_\xi(\bomega)^T\vdot\vv = \JJ_{\Gamma}(\zzt_0) \vdot \vv
	\label{eq:jvp_step_3}
\end{align}
which is the desired \jvp{} at $\zzt_0$.
To summarize, we can calculate the \jvp{} for the Jacobian $\JJ_{\Gamma}(\zzt_0)$
and an arbitrary vector $\vv\in\Reals^{2n}$ by first calculating the 
transformation $\xi(\bomega) = \JJ_{\Gamma}(\zzt_0)^T \vdot \bomega$ 
for any $\bomega\in \Reals^m$ (say $\omega_j = 1$ for all $j$), 
then calculating the \jtvp{} for this transformation $\xi$. 

Finally, we need a mechanism to calculate a \hvp{} 
of the form $\grad_\zeta^2 \ell(\bzeta) \vdot \vec{x}$ 
for some $\vec{x}\in \Reals^{2n}$. Noting that the Hessian is defined elementwise as
\begin{align}
	\left[\grad_\zeta^2 \ell(\vec{h})\right]_{ij} &= \pdv{}{h_i}\left[\grad_\zeta 
	\ell(\bzeta)\right]_j
	\label{eq:hessian_elementwise}
\end{align}
we can calculate $\grad_\zeta^2 \ell(\bzeta) \vdot \vec{x}$ 
elementwise as
\begin{align}
	\left[\grad_\zeta^2 \ell(\bzeta) \vdot \vec{x}\right]_i 
	= \sumj \left[\grad_\zeta^2 \ell(\bzeta)\right]_{ij} x_j
	= \sumj \pdv{}{h_i} \left[\grad_\zeta \ell (\bzeta)\right]_j x_j
	= \pdv{}{h_i}\left(\grad_\zeta \ell(\bzeta) \vdot \vec{x}\right).
	\label{eq:hvp_elementwise}
\end{align}
The full \hvp{} is therefore just the gradient $\grad_\zeta\left[\left(\grad_\zeta \ell(\hh) 
\vdot \vec{x}\right)\right]$, which is again a matrix-free procedure. 
This completes the desired 
mechanism for a matrix-free calculation of the GGN-vector-product. 

In practice, we formulate the LM linear system through the following sequence of calculations:
\begin{enumerate}
	\item We first calculate the \jtvp{} $\JJ(\zzt)^T \vdot 	\grad_{\zeta}\ell(\bzeta)$ 
	through the procedure in \Cref{eq:arbitrary_jtvp_sequence}. 
	This is just the gradient $\grad 
	f(\zzt)$, and can be accessed directly in any reverse-mode AD toolset.
	\item We set $\bomega = \grad_{\zeta}\ell(\bzeta)$, and thus $\xi(\zzt, \bomega) =  
	\JJ(\zzt)^T \vdot 	\grad_{\zeta}\ell(\bzeta)$, in 
	\cref{eq:jvp_step_1,eq:jvp_step_2,eq:jvp_step_3} to calculate the \jvp{} $\JJ(\zzt) \vdot 
	\Delta \zzt_\star$. While we can use any $\bomega\in\Reals^m$ for this calculation, 
	reusing the output from step (1) reduces the computational cost and memory overheads.
	\item We then calculate the \hvp{} 
	$\grad^2_{\bzeta} \ell(\bzeta) \vdot ( \JJ(\zzt) \vdot \Delta \zzt_\star)$ either: (i) 
	through the \hvp{} procedure described in \Cref{eq:hvp_elementwise} 
	or, (ii) if the Hessian $\grad^2_{\bzeta} \ell(\bzeta)$ is a diagonal matrix (and 
	therefore is easy 
	to formulate explicitly) then directly using the closed-form expression for 
	$\grad^2_{\bzeta} \ell(\bzeta)$. In case (ii), which is true for both the Gaussian and 
	Poisson error metrics, we only need to  calculate the $m$ diagonal elements of  
	$\grad^2_{\bzeta} \ell(\bzeta)$, then perform an elementwise product operation with the 
	vector $( \JJ(\zzt) \vdot \Delta \zzt_\star)$ to calculate the desired \hvp{}.
	\item Finally, we calculate the \jtvp{} 
	$\JJ^T (\zzt) \vdot ( \grad^2_{\bzeta} \ell(\bzeta) \vdot ( \JJ(\zzt) \vdot \Delta 
	\zzt_\star))$ through the procedure in \Cref{eq:arbitrary_jtvp_sequence}. This is again 
	easy to access from the AD toolset.
\end{enumerate} 
From step (1) and step (4) (and some extra elementary operations), we have the right-hand-side 
and left-hand-side of \Cref{eq:ggn_magnitude_appendix_2} respectively, which completes the 
full LM linear system.

\section{Calculating the diagonal elements of the Generalized Gauss-Newton matrices}
\label{appendix:jacobi_lm}

In this appendix, we derive general analytical expressions to calculate the diagonal
elements of the GGN matrices (defined in \Cref{eq:ggn}) for the ptychographic reconstruction
problems. For ease of analysis, 
this appendix defines the error metric $f$ as a function of the variable $\zz\in\Complexes^n$ 
and its complex conjugate $\zz^*$. This $(\zz, \zz^*)$ coordinate representation is connected 
to the
$(\Re{\zz}, \Im{\zz})$ representation used in \Cref{sec:LM_principles} through the 
Wirtinger gradient definition of
\begin{align}
	\grad f(\zz) = \pdv{f(\zz)}{\zz^*} = \frac{1}{2}\left(\pdv{f(\zz)}{\Rel{\zz}} 
	+ i\pdv{f(\zz)}{\Img{\zz}}\right),
	\label{eq:wirtinger}
\end{align}
with $i = \sqrt{-1}$, $\pdv*{f(\zz)}{\zz^*}$ the componentwise partial
derivative of $f$ with respect to the complex conjugate variable
$\zz^*$, and $\pdv*{f(\zz)}{\Rel{\zz}}$ and $\pdv*{f(\zz)}{\Img{\zz}}$
the componentwise partial derivatives with respect to the real and
imaginary parts of $\zz$. A change of representation is easily
accomplished \via{} the linear transform
\cite{sorber_siam_2012} of
\begin{align}
	\begin{bmatrix}
		\pdv*{f(\zz)}{\Rel{\zz}}\\
		\pdv*{f(\zz)}{\Img{\zz}}\end{bmatrix}
	= 
	\begin{bmatrix}
		\pdv*{f(\zz)}{\zz}\\
		\pdv*{f(\zz)}{\zz^*}
	\end{bmatrix}^T \vdot 
	\begin{bmatrix}
		\II_{n\times n}  & i\II_{n\times n}\\
		\II_{n\times n} & -i \II_{n\times n}
	\end{bmatrix}	 
\end{align}
where $\II_{n\times n}$ is the $n\times n$ identity matrix and $\cdot^T$ is
the transpose operator. 

For the SPR problem, our coordinates of interest are just $(\zz,\zz^*)=(\OO,\OOc)$.
For the magnitude-based formulation of the error metrics [\Cref{eq:error_metric}, 
\Cref{eq:error_metric_magnitude}], we 
can use the chain rule of multivariable calculus to 
derive
\begin{align}
	\grado f = \left(\pdv{\bzeta}{\OOc}\right)^T \vdot \grad{\ell}.
	\label{eq:grad_f_obj}
\end{align}
where  $\pdv{\bzeta}{\OOc}$ is the $(KM\times N)$ Jacobian matrix defined elementwise as
$\left[\pdv{\bzeta}{\OOc}\right]_{jn} = \frac{\partial
	\bzeta_j}{\partial \OOc_n}$. We can follow the procedure
in \cite{qian_ipa_2014} to calculate the set of $M$ lines in the
Jacobian matrix corresponding the $k$-th probe position.
Let $\bzeta_k := \{\zeta_j \,|\, j=(k-1)M, \cdots, kM\}$, so that the
sub-matrix in the Jacobian corresponding to the $k$-th
position is
\begin{align}
	\begin{split}
		\left[\pdv{\bzeta}{\OOc}\right]_k 
		&= \pdv{\bzeta_k}{\OOc}= \pdv{\hh_k^{1/2}}{\hh_k}\vdot\pdv{\hh_k}{\OOc}\\
		&= \diag{\frac{1}{2\hh_k^{1/2}}} \vdot \pdv{|\hbpsi_k|^2}{\OOc}\\
		&= \diag{\frac{1}{2\hh_k^{1/2}}}\vdot \pdv{\diag{\hbpsi_k}\vdot\hbpsic_k}{\OOc} \\
		&= \diag{\frac{1}{2\hh_k^{1/2}}}\vdot \left(\diag{\hbpsi_k}\vdot \pdv{\hbpsic_k}{\OOc} 
		+ 
		\left(\pdv{\diag{\hbpsi_k}}{\OOc}\right)\vdot \hbpsic_k\right)\\
		\left[\text{using }\pdv{\hbpsic}{\OOc} = \left(\pdv{\hbpsi}{\OO}\right)^*\right]\quad
		&= \diag{\frac{\hbpsi_k}{2\hh_k^{1/2}}}\vdot 
		\left(\pdv{\FF\vdot\diag{\PP}\vdot \SSk\vdot\OO}{\OO}\right)^*\\
		&= \diag{\frac{\hbpsi_k}{2\hh_k^{1/2}}}\vdot 
		\FF^*\vdot \diag{\PPc}  \vdot \SSk^* .
	\end{split}
	\label{eq:jac_ampl}
\end{align}
With this Jacobian in hand, the calculation of $\grado f$ (from \Cref{eq:grad_f_obj}) for a 
chosen error metric is straightforward, and we can now use $\grado f$ to formulate any choice 
of first-order algorithm. To formulate a second-order optimization algorithm, however, we need 
to additionally define
the ``compound'' Jacobian matrix $\JJc$ (see Eq. 116 in \cite{kreutz_arxiv_2009}) which 
contains the derivatives with respect to both the $\OO$ and $\OOc$ coordinates:
\begin{align}
	\JJc \coloneqq \left(\pdv{\bzeta}{\OO}\quad \pdv{\bzeta}{\OOc}\right) \in \Complexes^{KM 
		\times 2N}.
	\label{eq:jac_compund}
\end{align}
In other words, the matrices $\pdv*{\bzeta}{\OO}$ and $\pdv*{\bzeta}{\OOc}$ are stacked 
columnwise to form the matrix 
$\JJc$. Assuming that the error metric of interest, $\ell(\bzeta)$, has a positive 
semi-definite Hessian matrix defined 
elementwise as $\left[\hessl\right]_{k,j}=\frac{\partial^2 \ell}{\partial
	\zeta_k \partial \zeta_j}$, we can write the compound GGN matrix as
\begin{align}
	\GGlc = \JJc_\zeta^\dagger \vdot \hessl \vdot \JJc_\zeta 
	\label{eq:ggn_compound}
\end{align}
where $\bullet^\dagger$ is the conjugate transpose operation. From here on, we additionally 
assume 
that $\hessl$ is a diagonal 
matrix; this holds for both the Gaussian and Poisson 
error metrics.

Putting \Cref{eq:jac_compund} and \Cref{eq:ggn_compound} together, we get the relation
\begin{align}
	\GG_\OO = 
	\begin{pmatrix}
		(\pdv*{\bzeta}{\OO})^\dagger \vdot \hessl \vdot \pdv*{\bzeta}{\OO} & 
		(\pdv*{\bzeta}{\OO})^\dagger \vdot \hessl 
		\vdot \pdv*{\bzeta}{\OOc}\\
		(\pdv*{\bzeta}{\OOc})^\dagger \vdot \hessl \vdot \pdv*{\bzeta}{\OO} & 
		(\pdv*{\bzeta}{\OOc})^\dagger \vdot \hessl 
		\vdot \pdv*{\bzeta}{\OOc}
	\end{pmatrix}.
	\label{eq:ggn_compound_expanded}
\end{align}
We substitute for $\pdv{\bzeta}{\OOc}$ from \eqref{eq:jac_ampl}
to get the lower-right diagonal block in this latter matrix:
\begin{align*}
	\mat{D}
	=& \left(\pdv{\bzeta}{\OOc}\right)^\dagger \vdot \hessl \vdot \pdv{\bzeta}{\OOc}\\
	=& \sumk \left(\diag{\frac{\hbpsi_k}{2\hh_k^{1/2}}}\vdot \FF^*
	\vdot \diag{\PPc}  \vdot \SSk^*\right)^\dagger \vdot
	\grad^2_{k:} \ell\\
	\vdot \diag{\frac{\hbpsi_k}{2\hh_k^{1/2}}}\vdot \FF^* \vdot \diag{\PPc}  \vdot \SSk^*
\end{align*}
where $\grad^2_{k:} \ell$ is the $M\times M$ diagonal block extracted from
$\hessl$ that corresponds to the $k^{th}$ probe position. Since $\SSk$
is a binary matrix, we can use $\SSk^* = \SSk$ to get the simplified expression
\begin{align}
	\begin{split}
		\mat{D} &= \frac{1}{4}\sumk \SSk^T \vdot \diag{\PP} \vdot \mat{C}_k \vdot \diag{\PPc}  
		\vdot\SSk^* \quad \\ 
		\text{with}\quad \mat{C}_k &= \FF^T \vdot  \diag{\frac{\hbpsi_k^*}{\hh_k^{1/2}}} \vdot 
		\grad^2_{k:}\ell \vdot \diag{\frac{\hbpsi_k}{\hh_k^{1/2}}} \vdot \FF^*\\
		&\approx \FF^T \vdot \grad^2_{k:} \vdot \FF^*
	\end{split}  
	\label{eq:ggn_block_diag}
\end{align}
where we use the simplifying approximation\footnote{
	We only apply this approximation to calculate the diagonal scaling term $\bm{D}$ 
	and the preconditioner $\bm{P}$ (\Cref{subsec:lm_scaling_precond}).
	In our numerical tests, we do not see any appreciable difference in the LM iterates 
	obtained 
	with and without this approximation.}
$\hh_k\approx\abs{\hbpsi_k}^2$ and where $\mat{C}_k$ is a circulant matrix. The diagonal 
elements of 
$\mat{D}$ 
are then given by
\begin{align}
	d_n = \vec{e}_n^T \vdot \mat{D} \vdot \vec{e}_n \quad \text{for} \quad n=1\dots N,
	\label{eq:ggn_diag_1}
\end{align}
where $\vec{e}_n$ denotes a vector with a $1$ at index $n$ and $0$'s elsewhere. From 
\eqref{eq:ggn_block_diag} and 
\eqref{eq:ggn_diag_1} we have
\begin{align}
	d_n = \frac{1}{4}\sumk d_{k,n} \quad 
	\text{with}\quad 
	d_{k,n} = \vec{e}_n^T \vdot \SSk^T \vdot \diag{\PP} \vdot \mat{C}_k \vdot \diag{\PPc}  
	\vdot \SSk \vdot \vec{e}_n .
	\label{eq:ggn_diag_2}
\end{align}

Noting that $\SSk$ is a binary matrix that extracts the $M$ object pixels that interact with 
the probe at the $k^{th}$ 
probe position, we can see that
\begin{align*}
	\SSk \vdot \vec{e}_n = n^{th}\, \text{column of }\SSk = \vec{e}_{m'}\in\Reals^M
\end{align*}
where $\vec{e}_{m'}$ is a vector that contains all $0$'s except at index $m'$. The index $m'$ 
corresponds to the probe 
pixel that interacts with the object pixel $n$ at the $k^{th}$ probe position. Since there 
exist $N-M$ object pixels that 
do not interact with any probe pixel at a given probe position, $\vec{e}_{m'}$ contains a $1$ 
at index $m'$ if there 
exists such a probe pixel, and contains $0$'s everywhere otherwise. As a result, $\SSk \vdot 
\vec{e}_n$ is a vector that 
contains \textit{at most} a single $1$ among its elements. 
Then, for the diagonal matrix $ \diag{\PPc}$, the matrix-vector product 
\begin{align*}
	\vec{p}=\diag{\PPc}  \vdot \SSk \vdot \vec{e}_n
\end{align*} 
is also a vector that contains \textit{at most} a single non-zero element 
$\left[\PPc\right]_{m'}$ at position $m'$. 
If we then define the set $\Omega_n \subset \{1, 2, \cdots, K\}$ as the set of probe positions 
that interact with the 
sample pixel at index $n$, we get the relation (from \eqref{eq:ggn_diag_2})
\begin{align*}
	d_{k,n} = \vec{p}^\dagger \vdot \mat{C}_k \vdot \vec{p} 
	= \begin{cases}
		\left[\abs{\PP}^2\right]_{m'}\left[\mat{C}_k\right]_{m',m'} &\text{ if } k\in\Omega_n\\
		0 &\text{ if } k\notin\Omega_n.
	\end{cases}
\end{align*}
where $\left[\mat{C}_k\right]_{m',m'}$ is the element extracted from the position $m'$ in the 
main diagonal of 
$\mat{C}_k$. 

We now note that the circulant matrix $\mat{C}$ is constant along its
main diagonal and has the elements $[\mat{C}_k]_{m',m'} =
[\mat{C}_k]_{1,1}= \frac{1}{M}\mathtt{trace}\left(\grad^2_{k:} \ell\right)$. We finally have 
the relation
\begin{align}
	d_{k,n} 
	= \begin{cases}
		\frac{1}{M} \left[\abs{\PP}^2\right]_{m'}\mathtt{trace}\left(\grad^2_{k:} \ell\right) 
		&\text{ if } k\in\Omega_n\\
		0 &\text{ if } k\notin\Omega_n.
	\end{cases}
	\label{eq:ggn_diag_elem_final}
\end{align}
As an alternative, we can also interact directly with the shift matrices $\SSk$  instead of 
iterating over individual 
diagonal elements $d_n$. To accomplish this, we can note that
\begin{align*}
	\begin{split}
		d_{k,n} 
		&= \vec{e}_n^T \vdot \SSk^T \vdot \diag{\PP} \vdot \mat{C}_k \vdot \diag{\PPc}  \vdot 
		\SSk \vdot \vec{e}_n\\
		&= \vec{e}_n^T \vdot \SSk^T \vdot \diag{\PP} \vdot 
		\left(\frac{1}{M}\mathtt{trace}\left(\grad^2_{k:} \ell\right) \II_{M\times M}\right) 
		\vdot 
		\diag{\PPc}  \vdot \SSk \vdot \vec{e}_n
	\end{split}
\end{align*}
where $ \II_{M\times M}$ is the identity matrix. This gives us the relation
\begin{align}
	\begin{split}
		\Dg{\mat{D}} &= \Dg{\left(\pdv{\bzeta}{\OOc}\right)^\dagger \vdot \hessl \vdot 
		\pdv{\bzeta}{\OOc}}\\
		&= \frac{1}{4M} \sumk \mathtt{trace}\left(\grad^2_{k:} \ell\right) \Dg{ \SSk^T \vdot 
		\diag{\abs{\PP}^2} \vdot 
			\SSk}.
	\end{split}
	\label{eq:ggn_diag_mat}
\end{align}
If we follow the same procedure for the upper-left diagonal block in
\eqref{eq:ggn_compound_expanded}, we again get the same expression
as in \eqref{eq:ggn_diag_mat}. In practice, we can calculate these 
expressions in a straightforward manner by using array manipulation
tricks instead of using the matrix multiplication with the shift
$\SSk$. This holds even after we change the coordinate basis to
$(\Rel{}, \Img{})$ through the relationship (Equation 89
in \cite{kreutz_arxiv_2009})
\begin{align}
	[\GG_\OO]_{\Reals^{2N}} = 
	\begin{pmatrix}
		\II_{N\times N}  & i\II_{N\times N}\\
		\II_{N\times N} & -i \II_{N\times N}
	\end{pmatrix}^\dagger 
	\vdot \GG_\OO \vdot 
	\begin{pmatrix}
		\II_{N\times N}  & i\II_{N\times N}\\
		\II_{N\times N} & -i \II_{N\times N}
	\end{pmatrix}.
	\label{eq:ggn_coord_trans}
\end{align}

From our calculations thus far, we have obtained the desired analytical expressions 
we can use to calculate the diagonal elements of general GGN matrices for the SPR problem. 
For the BPR problem, we can follow a similar procedure to calculate the corresponding 
expression
for the probe variable. If we desire a joint object-probe optimization, the main
diagonal of the extended GGN matrix takes the form
\begin{align}
	\Dg{\GG_{\OO,\PP}} = 
	\begin{pmatrix}
		\Dg{\GG_\OO} 	& O\\
		0 						& \Dg{\GG_\PP}
	\end{pmatrix}
	\label{eq:ggn_joint_diag}
\end{align}
where $\GG_\PP$ is  the ``probe counterpart'' of the sample GGN matrix
defined by \eqref{eq:ggn_compound}.  We thus have the desired
analytical expressions for  the  diagonal elements of the GGN matrices
for both the SPR and BPR problems.

\section{Comparing magnitude-based and intensity-based GGN approximations}
\label{appendix:LM_magnitude_intensity}

In this appendix, we show that LM algorithm using magnitude-based GGN matrix ($\GG^\ell(\zzt)$ 
in \Cref{eq:ggn_magnitude}) makes for a faster optimization strategy than the scheme using 
the intensity-based GGN matrix ($\GG^\LL(\zzt)$). To accomplish this, we first examine the 
Gaussian error metric case, then develop  a guideline for how to formulate the GGN matrix for 
general error metrics. 

Using $\zzt=\OO$ for the SPR problem, from \Cref{eq:magnitude_nls,eq:grad_f_obj,eq:jac_ampl} 
we get an expression for the gradient of
\begin{align}
	\grad_{\OO} f_g &= \sumk \left(\FF^*\vdot \diag{\PPc}  \vdot \SSk^*\right)^T \vdot 
	\diag{\frac{\hbpsi_k}{2\hh_k^{1/2}}} \vdot \left(\hh_k^{1/2} - \yy_{k}^{1/2}\right)\\
	&= \frac{1}{2}\sumk \left(\FF^*\vdot \diag{\PPc}  \vdot \SSk^*\right)^T \vdot
	\diag{\hbpsi_k}\vdot \left(1 - \frac{\yy_{k}^{1/2}}{\hh_k^{1/2}}\right).
\end{align}

We are most concerned with the optimization iterations where $\zzt$ is in 
the neighborhood of the solution, since, for non-convex problems, this
is the region where 
second-order methods provide the most acceleration. In this region, with $h_{k,m}\in\hh_k$, 
$\hpsi_{k,m}\in \hbpsi_k$, $y_{k,m}\in\yy_k$ and $b_{k,m}\in \bb_k$, 
we get one of two cases: 
(i) $\abs{\hpsi_{k,m}}\gg b_{k,m}$, or 
(ii) all of $\abs{\hpsi_{k,m}}$, $b_{k,m}$, and $y_{k,m}$ have comparably small values, and 
therefore 
$|\hbpsi_{k,m} \left(1 - \tfrac{\yy_{k,m}^{1/2}}{\hh_{k,m}^{1/2}}\right)|$ is also very 
small in size and does not contribute meaningfully to the gradient calculation. For 
the data points of interest we can simply write 
\begin{align}
	h_{k,m} =\zeta^2_{k,m} \approx  \hpsi_{k,m}^2.
\end{align}
We can easily verify that a similar line of reasoning also holds for the Poisson error metric. 
We expect that this also holds in the general case.

We can now rewrite the expressions for $f^\LL$ and $f^\ell$ in 
\Cref{eq:error_metric_magnitude} as
\begin{align}
	&f = 
	\sumk \sum_{m=1}^M \LL(\,\cdot \,,\, y_{k,m}) \circ \overbrace{
		(\cdot)^2 \circ \abs{\hpsi_{k,m}}}^{\textrm{I}};
	&f^\ell = 
	\sumk \sum_{m=1}^M \ell(\,\cdot \,,\, y_{k,m}) \circ \overbrace{
		\abs{\hpsi_{k,m}}}^{\textrm{II}}.
	\label{eq:no_background}
\end{align}
Martens and Sutskever \cite{martens_icml_2011} have previously shown that if we compare the  
GGN matrix $\GG^\LL(\OO) = \JJ_{\textrm{I}}^T \vdot \grad^2_I \LL \vdot  \JJ_{\textrm{I}}$ 
with the true Hessian matrix $\grad^2_\OO f$,  the second derivative terms 
associated with $\LL$ are faithfully captured in $\GG^\LL(\OO)$. However, the second 
derivative terms associated with the calculations in (I) are not accurately captured in the 
GGN matrix.  Effectively, the more calculation we associate with the Jacobian terms in the GGN 
matrix, the more curvature information we lose. Since (I) contains an extra $(\cdot)^2$ 
operation in comparison with (II), we lose more curvature information when we calculate 
$\GG^\LL(\OO)$ than we do when we calculate $\GG^\ell(\OO)$. In other words, 
$\GG^\LL(\OO)$ is a less accurate approximation of $\grad^2_\OO f$ than is 
$\GG^\ell(\OO)$. 

This leads us to a general guideline: we should construct the 
functional decomposition of $f$ so that the function associated with the central Hessian part 
of the GGN matrix performs as much as possible of the computation in $f$ (see also 
\cite[Chapter 3.4]{martens_phd_2016}). 

\section{Efficacy of GGN optimizations for the Gaussian and Poisson error metrics}
\label{appendix:ggn_efficacy}

Examination of the individual GGN approximations for the Gaussian and
Poisson metrics to understand
the efficacy of GGN-based optimization procedures for these metrics. We first use 
\Cref{eq:ggn_approximation} to calculate the Hessian matrices
\begin{align}
	\label{eq:hessian_gaussian}
	\grad^2 \ffg(\zzt) \,&=\, \GGg(\bzeta) + 
	\mathfrak{S}_g, \quad \text{with} \quad 
	\mathfrak{S}_g = {\textstyle \sumj} (\zeta_j - y^{1/2}_j) \grad^2 \zeta_j \quad 
	\text{and}\\
	\label{eq:hessian_poisson}
	\grad^2 \ffp(\zzt) \,&=\, \GG_{p}(\zeta)  + 
	\mathfrak{S}_p, \quad \text{with} \quad  
	\mathfrak{S}_p = {\textstyle \sumj} (\zeta_j - y_j/\zeta_j) \grad^2 \zeta_j,
\end{align}
where $\GGg(\bzeta)$ and $\GGp(\bzeta)$ denote
the GGN matrices associated with the error metrics $\ellg$ and $\ellp$,
respectively, and $\mathfrak{S}_g$ and
$\mathfrak{S}_p$ denote the difference between the true
Hessian and the corresponding GGN approximation. When $\zzt$ is
far away from the minimum, we may get the extreme case with $\zeta_j
\ll 1$ and $y_j > 0$, where $\abs{\zeta_j - y_j/\zeta_j} >> \abs{\zeta_j - y^{1/2}_j}$. 
Due to contributions from these pixels, $\mathfrak{S}_p$ may
be much ``larger'' than $\mathfrak{S}_g$. This indicates
that $\GGp(\bzeta)$ is a less accurate approximation of
$\grad^2 \ffp(\zzt)$ than $\GGg(\bzeta)$ of
$\grad^2 \ffg(\zzt)$.

A second point is that the matrices $\GGp(\bzeta)$ and $\GGg(\bzeta)$ differ in the central 
Hessian term: $\grad^2 \ellg(\bzeta)$ is the identity matrix whereas $\grad^2 
\ellp(\bzeta)$ is the diagonal matrix with the elements $1 + y_j/\zeta_j^2$. If $\zeta_j \ll 
1$ and $y_j >0$, then the diagonal elements of $\grad^2 \ellp(\bzeta)$ 
can take very large values, which would degrade the conditioning of $\grad^2 \ellp(\bzeta)$ 
and, consequently, $\GGp(\bzeta)$. In fact, we can see from \Cref{eq:hessian_poisson} that 
this effect is present in general (Hessian-based) second-order optimization strategies, which 
makes the Gaussian error metric the preferred optimization target for general 
applications. 

Putting these effects together, we can see that $\ffp$ may be difficult to optimize 
through a GGN-based routine. 

\section{Defining a surrogate Poisson error metric}
\label{appendix:poisson_surrogate}

We can address the difficulty in optimizing the Poisson error metric with a GGN-based routine 
(see \Cref{appendix:ggn_efficacy}) by introducing a surrogate formulation that is 
asymptotically identical 
to the Poisson error metric we aim to minimize. We can build this surrogate formulation by 
adding a 
spatially uniform value to the incoherent background term in \Cref{eq:expected_count} and 
driving this value 
towards zero as we progress with the optimization iterations. For this procedure, we redefine 
the 
expected intensity $\hh_k$ (\Cref{eq:expected_count}) and magnitude $\bzeta_k$ 
(\Cref{eq:magnitude}) as:
\begin{align}
	\hh_k \,=\, \abs{\hbpsi_k}^2 + \bb_k + \varsigma_t; \qquad \bzeta_k \,=\, \hh_k^{1/2}
	\label{eq:expected_count_surrogate}
\end{align}
where $t$ indexes the optimization iteration and $0 \leq \varsigma_t < 1$. This redefinition 
leaves 
unchanged all the relations that define the preconditioner and the LM update.  We nevertheless 
need to 
define a constant $\varsigma_0$ and an integer $T\ge 1$ before we start the optimization 
procedure, 
then monotonically decrease the value of $\varsigma_t$ after every iteration until 
$t=T$, where we then set $\varsigma_T = 0$.  Therefore at $t=T$, the surrogate formulation is 
exactly equal 
to the Poisson error metric $\ellp$. Now, through a relatively large choice of $\varsigma_0$ 
(\eg{} $\varsigma=1$) we can significantly constrain the size of the elements in both the 
matrices $\mathfrak{S}_p$ and $\GGp(\bzeta)$ (which we define as in 
\Cref{appendix:ggn_efficacy}). We can expect this effect to become less significant as we 
progress get closer to the minimum, at which point we can transition to using the true Poisson 
error metric. 

For the numerical results presented in \Cref{sec:experiments}, we set $T=100$,  
$\varsigma_0=1$, and use an evenly spaced logarithmic grid to decrease the value of 
$\varsigma$ from $\varsigma_0 > 0$ to $0$.

\section{First-order optimization algorithms}
\label{appendix:first_order_algs}
In this appendix, we present the implementation details for the tested first-order algorithms. 
We use the reverse-mode AD procedure to calculate the gradients required within all the tested 
algorithms.

\subsection{ePIE}
The classic extended ptychographic engine (ePIE) algorithm 
\cite{maiden_ultramic_2009,maiden_optica_2017} optimizes the Gaussian error metric for the BPR 
problem by stochastically iterating through the individual diffraction patterns. 
In the ePIE method, the current probe and object estimates are updated concurrently as
\begin{align}
	\OO_{j+1} &= \OO_{j} - \alpha_{j} \delO \ffg(\OO_{j},\PP_{j}, \yy_k)\\
	\PP_{j+1} &= \PP_{j} - \gamma_{j} \delP \ffg(\OO_{j},\PP_{j}, \yy_k).
	\label{eq:epie}
\end{align}
Here, $\delO \ffg(\OO_j,\PP_j, \yy_k)$ and $\delP \ffg(\OO_j,\PP_j, \yy_k)$ are the 
derivatives of $\ffg$ with respect to $\OO$ and $\PP$ computed using only the information in 
the $k-$th diffraction pattern (chosen randomly). The step sizes 
$\alpha_j=1/{\maxnorm{\abs{\PP_j}^2}}$ and $\gamma_j=1/\maxnorm{\abs{\SSk\vdot\OO_j}^2}$ are 
the inverse of the Lipschitz constants of the partial gradients $\delO \ffg(\OO_j,\PP_j, 
\yy_k)$ and $\delP \ffg(\OO_j,\PP_j, \yy_k)$ respectively.

\subsection{Nonlinear conjugate gradient }

%

We use the popular Polak-Ribiere nonlinear conjugate gradient (NCG) method \cite[Chapter 
5]{nocedal_2006}, 
with the update step sizes calculated using an adaptive backtracking line search procedure,
to solve the unconstrained SPR problem for both the Gaussian and Poisson error 
metrics \cite{thibault_njp_2012}. 

For ill-conditioned problems, we can 
accelerate the basic NCG algorithm by choosing a matrix that approximates 
$\left[\grad^2 f\right]^{-1}$ as a preconditioner \cite{hager_pjo_2006}. 
Since the GGN matrix can be considered to be a proxy for the Hessian, 
the matrix $\Dg{\GG}^{-1}$ seems a sensible choice as a preconditioner 
for the NCG algorithm. We therefore modify the standard NCG method for each of the 
Gaussian and Poisson error metrics to use this preconditioner and denote 
this algorithm as the preconditioned nonlinear conjugate gradient (PNCG) method. 
In \Cref{sec:experiments,sec:discussion}, we only report the results with the PNCG algorithm,
but we report the full results for both the PNCG and NCG algorithms in 
\Cref{appendix:costs_comparison}.
In practice, we find that the use of this preconditioner provides only slight acceleration to 
the basic
NCG algorithm.

\subsection{Nesterov's accelerated gradient}
The Nesterov's accelerated gradient (NAG) method, also referred to as the 
Nesterov's momentum method, is an accelerated adaptation of the 
standard gradient descent method. The NAG method in the unconstrained 
SPR setting for the Gaussian error metric uses the scheme
\begin{align}
	\vec{v}_{j+1} &= \gamma_j v_j + \alpha \grad_{\OO} f_A (\OO_j)\\
	\OO_{j+1} &= \OO_j + \vec{v}_{j + 1}.
	\label{eq:nag}
\end{align}
Here, the ``velocity'' term $\vec{v}\in\Reals^{2N}$ 
stores a weighted history of the past gradient directions
and uses this to adapt the current update direction. 
For compatibility with existing phase retrieval literature, 
we set the step size parameter to
$\alpha=1/\maxnorm{\sum_{k=1}^K\abs{\SSk^\dagger \vdot \PP}^2}$, 
the inverse of the Lipshcitz constant of the gradient $\grad_{\OO} \ffg(\OO_j)$ 
\cite{hesse_siamjis_2015}. 
Finally, we use the standard fixed schedule $\gamma_j = (j+2)/(j+5)$ for the momentum 
parameter $\gamma_j$ \cite{sutskever_icasp_2013}.

If we set $\gamma_j=0$, 
the resulting algorithm is exactly the 
``reshaped Wirtinger flow'' algorithm \cite{zhang_anips_2016} 
only with a different the step size parameter.
The NAG algorithm is also closely related to the 
``accelerated Wirtinger flow'' algorithm proposed in \cite{xu_arxiv_2018}
that uses a scaled gradient in addition to Nesterov's momentum.

\subsection{Proximal heterogeneous block implicit-explicit method}
The proximal heterogeneous block implicit-explicit  (PHeBIE) 
scheme described in \cite{hesse_siamjis_2015} 
is provably convergent for the constrained BPR problem 
with the magnitude-based error metric. 
The PHeBIE scheme treats the transmitted waves 
$\bPsi=\{\bpsi_1,\dots,\bpsi_K\}$ as an auxiliary variable that
is kept fixed during the probe and object updates. The updates are calculated as
\begin{align}
	\OO_{j+1} &= \Pi_{\setO}\left(\OO_j - \alpha_{j+1} \grad_{\sOO} f_A (\OO_j, \PP_j, 
	\bPsi_j) \right)\\
	\PP_{j+1} &= \Pi_{\setP}\left(\PP_j - \gamma_{j+1} \grad_{\sPP} f_A (\OO_{j+1}, \PP_j, 
	\bPsi_j) \right)\\
	\bpsi_{k,j+1} &= \diag{\PP_{j+1}}\vdot \SSk \vdot \OO_{j+1} \quad \text{for }1\leq k \leq 
	K.
	\label{eq:phebie}
\end{align}
where $\Pi_\setO$ and $\Pi_\setP$ represent the projections into the convex sets $\setO$ and 
$\setP$ respectively.
The object and probe step sizes 
\begin{align}
	\alpha_{j+1} = 1/\maxnorm{\sum_{k=1}^K\abs{\SSk^\dagger \vdot \PP_j}^2} \quad \text{and} 
	\quad
	\gamma_{j+1} = 1/\maxnorm{\sum_{k=1}^K\abs{\SSk \vdot \OO_j}^2}
	\label{eq:phebie_step_sizes}
\end{align}
are again derived from the partial Lipschitz constants for the respective gradients.

\subsection{Alternating Directions Minimization}
\label{appendix:admm}
The Alternating Directions Minimization (ADMM) method 
\cite{birkhoff_tams_1959,birkhoff_advcomp_1962} has recently been adapted 
as a provably convergent scheme for the BPR problem for both the magnitude-based 
and Poisson error metrics \cite{chang_sjis_2019}.
To formulate the ADMM scheme, we have to first define the operator
\begin{align*}
	\bm{A}(\OO,\PP) = \begin{bmatrix}
		\FF\vdot \diag{\PP}\vdot \SSn_1 \vdot \OO\\
		\vdots\\
		\FF\vdot \diag{\PP}\vdot \SSn_K \vdot \OO
	\end{bmatrix}
\end{align*}
which generates the transmitted waves at the far-field detector plane. We can then define an 
auxiliary 
variable $\hbPsi=\bm{A}(\OO, \PP)$ that is kept fixed during the probe and object updates. The 
``augmented 
Lagrangian'' for the ADMM scheme is then given by
\begin{align}
	\Upsilon_\varrho(\OO, \PP, \hbPsi, \bLambda) = 
	f(\OO, \PP, \hbPsi) + \Rel{(\hbPsi - \bm{A}(\OO, \PP))\vdot \bLambda} 
	+ \frac{\varrho}{2}\norm{\hbPsi - \bm{A}(\OO, \PP)}^2
	\label{eq:admm_lagrangian}
\end{align}
with $\OO\in\setO$ and $\PP\in\setP$. Here, $f(\OO, \PP, \hbPsi)$ is the objective function 
(either Gaussian 
or Poisson), $\bLambda\in\Complexes^{KM}$ the Lagrange multiplier, and $\varrho>0$ the penalty 
parameter. The 
variable updates are as follows:
\begin{align}
	\PP_{j+1} 
	&= \argmin_{\PP\in\setP}\Upsilon_\varrho(\PP,\OO_j,\hbpsi_j,\bLambda_j) \\
	&= \Pi_{\setP}\left(\frac{\sumk \SSk \vdot \diag{\OO_j}^\dagger \vdot \FF^ \dagger\vdot 
	(\varrho\, \hbpsi_{k,j} + \Lambda_{k,j})}{\varrho \sumk \abs{\SSk\vdot \OO_j}^2}\right)\\
	\OO_{j+1} 
	&= \argmin_{\OO\in\setO}\Upsilon_\varrho(\PP_{j+1},\OO,\hbpsi_j,\bLambda_j)\\
	&=	\Pi_{\setO}\left(\frac{\sumk \SSk^\dagger \vdot \diag{\PP_{j+1}}^\dagger \vdot 
	\FF^\dagger \vdot (\varrho\, \hbpsi_{k,j} + \Lambda_{k,j})}{\varrho \sumk 
	\abs{\SSk^\dagger \vdot \PP_{j+1}}^2}\right)\\
	\hbPsi_{j+1} &= \argmin_{\hbPsi} \Upsilon_\varrho(\PP_{j+1},\OO_{j+1},\hbpsi,\bLambda_j)\\
	\bLambda_{j+1} &= \bLambda_j + \varrho \left(\hbPsi_{j+1} - 
	\bm{A}(\OO_{j+1},\PP_{j+1})\right).
	\label{eq:admm_updates}
\end{align}
The object and probe updates use the exact solutions to their respective
minimization problems. For the auxiliary variable update, we
we use a single iteration of the projected gradient algorithm 
to solve Equation 3.10 in \cite{chang_sjis_2019}.

\begin{figure}[th]
	\includegraphics[width=\linewidth]{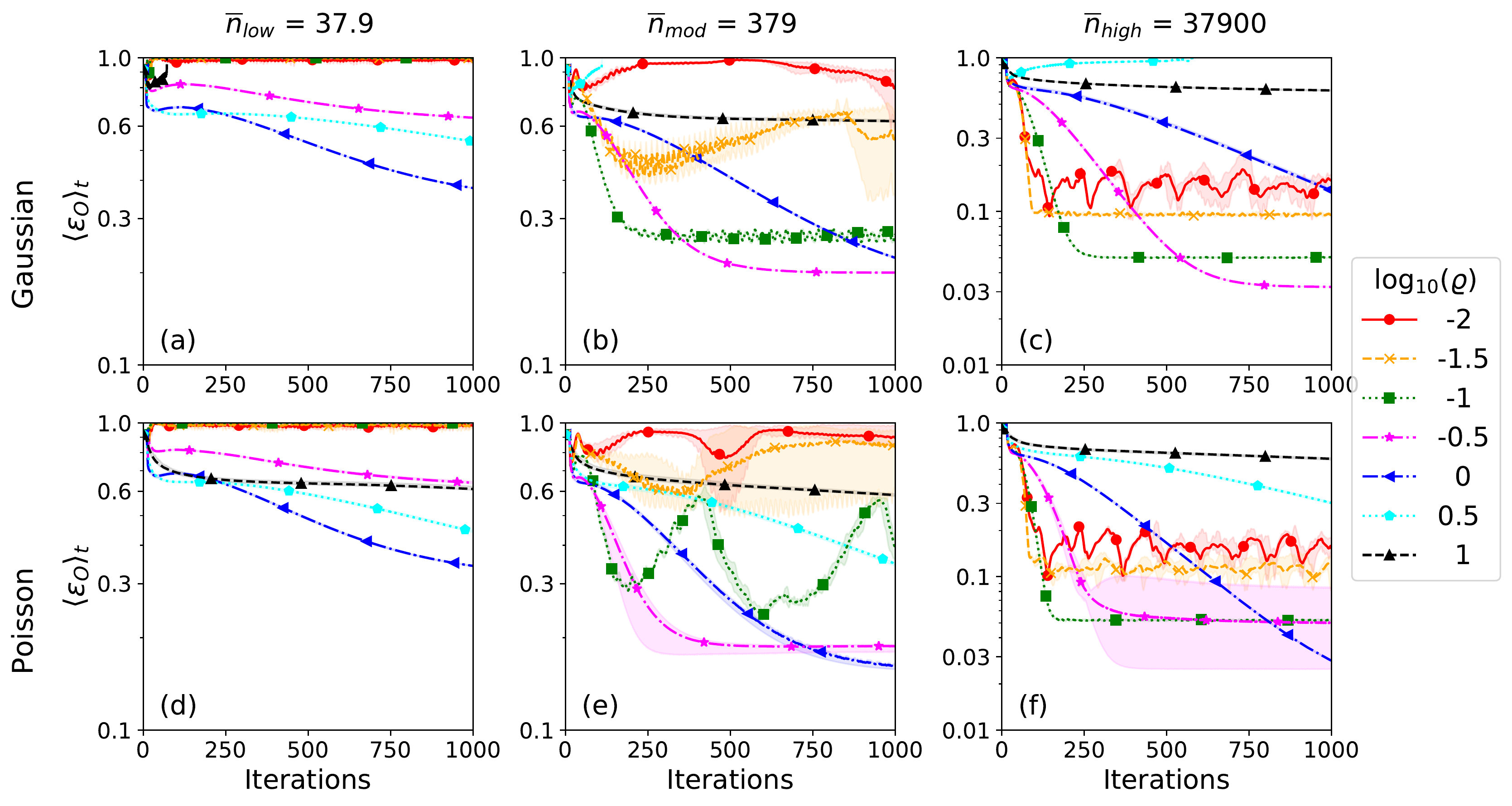}
	\caption{Effect on the normalized object reconstruction error \oerr{} as a function of the 
	$\varrho$ penalty parameter for the ADMM reconstructions. The left, mid, and right columns 
	display the plots generated for the \flow{}, \fmod{}, and \fhigh{}  settings respectively 
	for the Gaussian error metric (top), and the Poisson error metric (bottom).  For the 
	convergence analysis in \Cref{sec:experiments},
		and \Cref{appendix:costs_comparison} we use (a) $\varrho=1.0$,  $\varrho=10^{-0.5}$, 
		(b) $\varrho=10^{-0.5}$,
		(c) $\varrho=0.1$, (d) $\varrho=1.0$, (e) $\varrho=1.0$, and (f) $\varrho=1.0$.
	}
	\label{fig:admm_varrho}
\end{figure}

The performance of the ADMM method depends strongly on the choice of the penalty parameter 
$\varrho$ \cite{boyd_2011}. However, we are not aware of any practical guideline on how to 
choose the optimal value of $\varrho$ for the BPR optimization problem. 
In our experiments, we find that the large values of $\varrho$ obtained by following the proof 
strategy
in \cite{chang_sjis_2019} lead to impractically slow optimization; the numerical experiments 
in \cite{chang_sjis_2019} instead use much smaller, manually tuned, values of $\varrho$. Thus, 
to find the optimal 
value of the penalty parameter $\varrho$,
we run separate reconstructions for $\varrho=10^x$ for $x\in\{-2,\, -1.5,\, -1,\, -0.5,\, 0,\, 
0.5,\, 1\}$
for each simulation setting and error metric. 
For the convergence analysis in this work, we choose the value of $\varrho$ that enables a 
monotonic descent of the objective function $f$. Additionally, when multiple choices of 
$\varrho$ yield similar final values of $f$ (after 1000 iterations), we choose the value of 
$\varrho$ that leads to the lowest object reconstruction error $\oerr_t$. While this latter 
choice is  \textit{post-hoc} in nature and is impractical for actual ptychography experiments, 
it suffices for our analysis. We present, in \Cref{fig:admm_varrho}, the convergence history 
of reconstructions as a function of
the $\varrho$ parameter for the various numerical experiments we report in this paper.

Finally, we note that, first, our tuning procedure is computationally demanding and results in 
choices 
of $\varrho$ that are much smaller than that used in the proof strategy in 
\cite{chang_sjis_2019}, and therefore do not guarantee the convergence of the object 
and probe updates. Second, it is likely that a finer tuning of $\varrho$ could result in a 
more performant ADMM procedure, but this would require significant additional computational 
effort and is thus beyond the scope of this paper.

\section{Computational costs for the compared algorithms}
\label{appendix:costs_comparison}

\setlength{\tabcolsep}{3pt}
\renewcommand{\arraystretch}{1.0}

In this appendix, we report the final reconstruction results for all the numerical experiments 
we report in this work.
\subsection{Reconstruction results for the SPR experiments}
\label{sec:spr_results_description}

In \Cref{tab:gaussian_spr}, we report the final reconstruction results obtained by optimizing 
the Gaussian error metric for the SPR problem for all the three tested SNR scenarios. 
In the table, ``$It.$'' denotes the 
number of outer iterations required for convergence, and 
``$I_\OO$'' denotes the number of inner iterations required. Specifically, the ``inner 
iterations'' 
reported for the LM algorithms uses the format ``$\cdot/\cdot/\cdot$'' and contains the total
number of CG iterations, projected gradient calls (outer), 
and projected gradient line search iterations (inner) respectively. For the NCG algorithms, 
``$I_\OO$'' is just the total number of line searches iterations required. 

For the results reported, the points of convergence were calculated using 
\Cref{eq:window_rmsd_conv} with $\epsilon_c=3\times10^{-3}$, $\epsilon_c=2\times 10^{-3}$, and 
$\epsilon_c=10^{-3}$ respectively for the \flow{}, \fmod{}, and \fhigh{} cases.
\begin{table}
	\small
	\centering
	\begin{tabular}{ l c c c c c@{}} 
		\toprule
		Alg. & $\langle It.\rangle$ & $\langle I_\OO \rangle$ & $\langle f_g\rangle$ & \oerr{} 
		& \vtop{\hbox{\strut Flops}\hbox{\strut $\times10^{11}$}}\\ 
		\toprule
		\multicolumn{6}{@{}l}{\textbf{Low SNR }(\flow{})}\\
		PLM       &       20 &   86/0/0 & 6.88e+04 &     0.24 &        3 \\
		LM        &       10 &   63/0/0 & 6.92e+04 &     0.25 &        2 \\
		NAG       &      152 &        - & 6.88e+04 &     0.24 &        2 \\
		PNCG      &       68 &      111 & 6.88e+04 &     0.25 &        3 \\
		NCG       &       83 &      142 & 6.88e+04 &     0.25 &        3 \\
		\midrule
		\multicolumn{6}{@{}l}{\textbf{Moderate SNR }(\fmod{})}\\
		PLM       &       13 &   97/0/0 & 1.32e+05 &     0.15 &        3 \\
		LM        &       12 &   93/0/0 & 1.32e+05 &     0.14 &        3 \\
		NAG       &      216 &        - & 1.31e+05 &     0.13 &        3 \\
		PNCG      &       74 &      120 & 1.31e+05 &     0.14 &        3 \\
		NCG       &       82 &      145 & 1.31e+05 &     0.14 &        3 \\
		\midrule
		\multicolumn{6}{@{}l}{\textbf{High SNR }(\fhigh{})}\\
		PLM       &       19 &  110/0/0 & 5.81e+05 &    0.023 &        4 \\
		LM        &       17 &  144/0/0 & 5.85e+05 &    0.024 &        4 \\
		NAG       &      381 &        - &  5.7e+05 &    0.021 &        6 \\
		PNCG      &       96 &      158 & 5.71e+05 &    0.023 &        4 \\
		NCG       &       93 &      163 & 5.72e+05 &    0.022 &        4 \\
		\bottomrule
	\end{tabular}
	\label{tab:gaussian_spr}
	\caption{Reconstruction results and computational costs at the point of convergence for 
	the SPR problem with the Gaussian error metric.}
\end{table}
Table \Cref{tab:poisson_spr} also reports the final reconstruction results obtained by 
optimizing the 
Poisson error metric for the SPR problem. 
In this case, the ``PLM-S'' algorithm reported is the LM algorithm  optimizing the surrogate 
formulation  of the Poisson error metric (\Cref{appendix:poisson_surrogate}).
\begin{table}
	\small
	\centering
	\begin{tabular}{ l c c c c c@{}} 
		\toprule
		Alg. & $\langle It.\rangle$ & $\langle I_\OO \rangle$ & $\langle f_p\rangle$ & \oerr{} 
		&\vtop{\hbox{\strut Flops}\hbox{\strut $\times10^{11}$}}\\ 
		\toprule
		\multicolumn{6}{@{}l}{\textbf{Low SNR }(\flow{})}\\
		PLM       &       36 &  154/0/0 &  1.5e+05 &     0.24 &        6 \\
		LM        &       42 &  218/0/0 &  1.5e+05 &     0.23 &        7 \\
		PLM-S     &       49 &  135/0/0 &  1.5e+05 &     0.24 &        6 \\
		LM-S      &       72 &  156/0/0 &  1.5e+05 &     0.23 &        7 \\
		PNCG      &       93 &      159 &  1.5e+05 &     0.23 &        4 \\
		NCG       &       90 &      156 &  1.5e+05 &     0.24 &        4 \\
		\midrule
		\multicolumn{6}{@{}l}{\textbf{Moderate SNR }(\fmod{})}\\
		PLM       &       73 &  344/0/0 & -3.52e+06 &     0.11 &       12 \\
		LM        &       74 &  543/0/0 & -3.52e+06 &     0.11 &       17 \\
		PLM-S     &       51 &   79/0/0 & -3.52e+06 &     0.11 &        4 \\
		LM-S      &       56 &  110/0/0 & -3.52e+06 &     0.11 &        5 \\
		PNCG      &       87 &      150 & -3.52e+06 &     0.11 &        4 \\
		NCG       &      107 &      190 & -3.52e+06 &     0.11 &        5 \\
		\midrule
		\multicolumn{6}{@{}l}{\textbf{High SNR }(\fhigh{})}\\
		PLM       &       99 &  1385/0/0 & -1.36e+09 &    0.026 &       38 \\
		LM        &      109 &  1932/0/0 & -1.36e+09 &    0.031 &       52 \\
		PLM-S     &       74 &   221/0/0 & -1.36e+09 &    0.029 &        9 \\
		LM-S      &       93 &   273/0/0 & -1.36e+09 &    0.027 &       11 \\
		PNCG      &      125 &       217 & -1.36e+09 &    0.017 &        5 \\
		NCG       &      131 &       233 & -1.36e+09 &    0.023 &        6 \\
		\bottomrule
	\end{tabular}
	\label{tab:poisson_spr}
	\caption{Reconstruction results and computational costs at the point of convergence for 
	the SPR problem with the Poisson error metric.}
\end{table}
\subsection{Reconstruction results for the BPR experiments}
\label{sec:bpr_results_description}
Tables \Cref{tab:gaussian_bpr,tab:poisson_bpr} contain the reconstruction results obtained by 
optimizing 
the Gaussian error metric and Poisson error metrics
for the BPR problem. 
In this case, for the LM-A and NCG algorithms, 
``$I_\OO$'' and ``$I_\PP$'' denote the inner iterations 
required for the object and probe updates respectively. 
For the PLM-J algorithm, the object and probe updates are jointly calculated, 
but, for convenience, we still use the ``$I_\OO$'' column to report the number of 
inner iterations required. Similarly, for convenience, in the ADMM case, ``$I_\OO$'' denotes 
the number of line search iterations within the projected gradient algorithm.
Finally, $\epsilon_\PP$ denotes the normalized error 
for the reconstructed probe variable at the point of convergence. As with 
\Cref{sec:spr_results_description}, the points of convergence were calculated using 
\Cref{eq:window_rmsd_conv} with $\epsilon_c=3\times10^{-3}$, $\epsilon_c=2\times 10^{-3}$, and 
$\epsilon_c=10^{-3}$ respectively for the \flow{}, \fmod{}, and \fhigh{} cases.
\begin{table}
	\small
	\centering
	\begin{tabular}{ l c c c c c c c@{}} 
		\toprule
		Alg. & $\langle It.\rangle$ & $\langle I_\OO \rangle$ & $\langle I_\PP \rangle$ & 
		$\langle f_g\rangle$ & \oerr{} 
		& $\perr$ & \vtop{\hbox{\strut Flops}\hbox{\strut $\times10^{11}$}}\\ 
		\toprule
		\multicolumn{8}{@{}l}{\textbf{Low SNR }(\flow{})}\\
		PLM-A     &       29 &  97/5/10 &   76/0/0 & 6.77e+04 &     0.25 &    0.082 &        6 
		\\
		LM-A      &       12 &   68/0/0 &   42/0/0 & 6.79e+04 &     0.26 &    0.087 &        3 
		\\
		PLM-J     &       15 &   82/2/2 &        - & 6.81e+04 &     0.24 &    0.095 &        3 
		\\
		ADMM      &      999 &      1494 &        - &    7e+04 &     0.38 &     0.11 &       
		28 \\
		PHeBIE    &      868 &         - &        - & 6.77e+04 &     0.26 &    0.083 &       
		24 \\
		ePIE      &      267 &         - &        - & 1.03e+06 &     0.84 &     0.44 &        
		6 \\
		\midrule
		\multicolumn{8}{@{}l}{\textbf{Moderate SNR }(\fmod{})}\\
		PLM-A     &       14 &   77/0/0 &   42/0/0 &  1.3e+05 &     0.14 &    0.045 &        4 
		\\
		LM-A      &       13 &   77/0/0 &   46/0/0 & 1.31e+05 &     0.14 &    0.044 &        4 
		\\
		PLM-J     &       14 &   98/0/0 &        - & 1.31e+05 &     0.14 &    0.058 &        3 
		\\
		ADMM      &      640 &      954 &        - & 1.39e+05 &      0.2 &    0.042 &       18 
		\\
		PHeBIE    &      999 &        - &        - &  1.3e+05 &     0.16 &    0.037 &       28 
		\\
		ePIE      &      999 &        - &        - & 1.63e+05 &     0.22 &     0.17 &       21 
		\\
		\midrule
		\multicolumn{8}{@{}l}{\textbf{High SNR }(\fhigh{})}\\
		PLM-A     &       63 &  323/0/2 &  103/0/0 & 5.69e+05 &    0.021 &    0.012 &       16 
		\\
		LM-A      &       67 &  403/0/0 &  111/0/0 & 5.69e+05 &    0.021 &    0.012 &       19 
		\\
		PLM-J     &       22 &  189/0/0 &        - & 5.85e+05 &    0.026 &   0.0095 &        6 
		\\
		ADMM      &      731 &     1089 &        - & 6.04e+05 &    0.034 &   0.0051 &       21 
		\\
		PHeBIE    &      999 &        - &        - & 5.86e+05 &     0.08 &    0.011 &       28 
		\\
		ePIE      &      999 &        - &        - & 6.72e+05 &    0.074 &    0.044 &       21 
		\\
		\bottomrule
	\end{tabular}
	\label{tab:gaussian_bpr}
	\caption{Reconstruction results and computational costs at the point of convergence for the
		BPR problem with the Gaussian error metric.}
\end{table}
\begin{table}
	\small
	\centering
	\begin{tabular}{ l c c c c c c c@{}} 
		\toprule
		Alg. & $\langle It.\rangle$ & $\langle I_\OO \rangle$ & $\langle I_\PP \rangle$ & 
		$\langle f_g\rangle$ & \oerr{} 
		& $\perr$ & \vtop{\hbox{\strut Flops}\hbox{\strut $\times10^{11}$}}\\ 
		\toprule
		\multicolumn{8}{@{}l}{\textbf{Low SNR}}\\
		PLM-A     &       81 &     152/5/10 &   90/0/0 & 1.49e+05 &     0.23 &    0.081 
		&       15 \\
		LM-A      &       78 &     284/6/13 &  101/0/0 & 1.49e+05 &     0.23 &     0.08 
		&       18 \\
		PLM-A-S   &      113 &    110/22/39 &  106/0/0 & 1.49e+05 &     0.22 &    0.073 
		&       19 \\
		LM-A-S    &      121 &    126/16/30 &  117/0/0 & 1.49e+05 &     0.22 &    0.076 
		&       20 \\
		PLM-J     &      128 &    243/27/48 &        - & 1.49e+05 &     0.23 &    0.077 
		&       16 \\
		PLM-J-S   &      280 &  403/128/229 &        - & 1.49e+05 &     0.25 &    0.076 
		&       34 \\
		ADMM      &      999 &         1493 &        - & 2.05e+05 &     0.34 &    0.094 
		&       29 \\
		\midrule
		\multicolumn{8}{@{}l}{\textbf{Moderate SNR}}\\
		PLM-A     &       90 &    305/5/9 &  162/0/0 & -3.52e+06 &     0.11 &    0.035 &       
		20 \\
		LM-A      &       91 &    531/4/7 &  171/0/0 & -3.52e+06 &     0.11 &    0.036 &       
		25 \\
		PLM-A-S   &       88 &    94/5/11 &   76/0/0 & -3.52e+06 &     0.11 &    0.039 &       
		15 \\
		LM-A-S    &      100 &   130/5/12 &   96/0/0 & -3.52e+06 &     0.11 &    0.043 &       
		17 \\
		PLM-J     &      133 &   423/8/13 &        - & -3.52e+06 &     0.11 &     0.04 &       
		21 \\
		PLM-J-S   &      169 &  162/15/28 &        - & -3.52e+06 &      0.1 &    0.037 &       
		17 \\
		ADMM      &      866 &       1372 &        - &  -3.4e+06 &     0.17 &    0.033 &       
		25 \\
		\midrule
		\multicolumn{8}{@{}l}{\textbf{High SNR}}\\
		PLM-A     &       95 &  1133/21/34 &  477/0/0 & -1.36e+09 &    0.027 &     0.02 
		&       41 \\
		LM-A      &       99 &  1765/22/36 &  537/0/0 & -1.36e+09 &    0.028 &     0.02 
		&       56 \\
		PLM-A-S   &      120 &   323/37/62 &  143/0/0 & -1.36e+09 &    0.028 &    0.016 
		&       25 \\
		LM-A-S    &      102 &   335/16/27 &  129/0/0 & -1.36e+09 &    0.037 &    0.019 
		&       22 \\
		PLM-J     &      112 &  1275/22/37 &        - & -1.36e+09 &     0.02 &   0.0087 
		&       40 \\
		PLM-J-S   &      141 &   579/30/47 &        - & -1.36e+09 &    0.024 &    0.016 
		&       26 \\
		ADMM      &      999 &        1491 &        - & -1.36e+09 &    0.028 &    0.011 
		&       29 \\
		\bottomrule
	\end{tabular}
	\label{tab:poisson_bpr}
	\caption{Reconstruction results and computational costs at the point of convergence for the
		BPR problem with the Poisson error metric.}
\end{table}

\clearpage
\bibliography{../xrmbook}
\bibliographystyle{jabbrv_ieeetr}

\end{document}